\documentclass[aps,prd,twocolumn,showpacs,nofootinbib,amsmath,amssymb,amsfonts,showkeys]{revtex4}
\usepackage{graphicx}
\usepackage{bm}

\begin{document}
\title{Sound speed of scalar field dark energy: \\ weak effects and large uncertainties}

\author{Olga Sergijenko}
 \email{olka@astro.franko.lviv.ua}
\affiliation{Astronomical Observatory of 
Ivan Franko National University of Lviv, Kyryla i Methodia str., 8, Lviv, 79005, Ukraine\\
Main Astronomical Observatory of the National Academy of Sciences of Ukraine, Zabolotnoho str., 27, Kyiv, 03680, Ukraine}
\author{Bohdan Novosyadlyj}
 \email{novos@astro.franko.lviv.ua}
\affiliation{Astronomical Observatory of 
Ivan Franko National University of Lviv, Kyryla i Methodia str., 8, Lviv, 79005, Ukraine\\
Ya. S. Pidstryhach Institute for Applied Problems of Mechanics and Mathematics, Naukova str., 3-b, Lviv, 79060, Ukraine}

\date{\today}

\begin{abstract} 
The possibility of reconstruction of Lagrangian for the scalar field dark energy with constant effective sound speed $c_s$ is analyzed. It is found that such reconstruction can be made with accuracy up to an arbitrary constant. The value of $c_s$ is estimated together with other dark energy parameters ($\Omega_{de}$, $w_0$, $c_a^2$) and main cosmological ones on the basis of data including Planck-2013 results on CMB anisotropy, BAO distance ratios from recent galaxy surveys, galaxy power spectrum from WiggleZ, magnitude-redshift relations for distant SNe Ia from SNLS3 and Union2.1 compilations, the HST determination of the Hubble constant. It is shown that no value of $c_s$ from the range [0,1] is preferred by the used data because of very weak influence of dark energy perturbations on the large scale structure formation and CMB temperature fluctuations.
\end{abstract}
\maketitle

\section{Introduction}

Extragalactic and cosmological observational data collected up to now and interpreted in the framework of current physical theories certify that about 70\% of the energy-mass content of our World is the dark energy which fills the Universe almost uniformly and accelerates its expansion. Its physical nature is still unknown because of its ``darkness'' and exclusively cosmological scale ``fingerprints''. The explanation of the nature of this mysterious component becomes extremely important for elaboration of physics of galaxies and clusters, cosmology and particle physics beyond the Standard Model.

Among several discussed in the literature hypotheses about the nature of dark energy the hypothesis that it is a scalar field with violated weak or null energy condition seems the most promising in the terms of possibility to be tested by comparison of theoretical predictions with observational data. The scalar field can be quintessential dark energy, phantom one or changing the type from one to another (quintom) at different moments of time, or be the vacuum-like (or $\Lambda$-type) dark energy. The current observational data on supernovae type Ia (SNe Ia) luminosity distances, baryon acoustic oscillations (BAO) in the galaxies space distribution together with the Planck measurements of CMB temperature anisotropy \cite{Planck2013a,Planck2013b,Planck2013c} prefer phantom dark energy at 2$\sigma$ or a bit higher confidential level \cite{Xia2013,Rest2013,Cheng2013,Shafer2013,Novosyadlyj2014}. The accuracy of obtained in recent years observational data has increased so much that reliable determination of the 
equation of state parameter of dark energy $w_0$ (with accuracy of ~6\%) and its density parameter $\Omega_{de}$ (with accuracy ~2\%) for current epoch became possible. Moreover, the nowadays observational data are so good that they even give the possibility to establish also the time variation of $w_{de}$ (more precisely, the squared adiabatic sound speed $c_a^2\equiv\dot{p}_{de}/\dot{\rho}_{de}$ with accuracy ~18\% \cite{Novosyadlyj2014}). So, its reliable determination may be a matter of expected data in the nearest future. These parameters, however, are not enough to establish definitively the nature of dark energy. Other measurements of dark energy must be done.  
The determination of effective sound speed of dark energy\footnote{The terms ``adiabatic sound speed'' and ``effective sound speed'' of dark energy are used in the literature for designation of dark energy values which formally correspond to the thermodynamical ones.} $c_s$, which is the speed of propagation of dark energy perturbations, is among them.

The theoretical aspects of effective sound speed of dark energy, the impact of its value on the evolution of dark energy and dark matter perturbations, the possibility of its determination as well as the recent attempts of its constraining are analyzed in \cite{Kodama1984,Hu1998,Erickson2002,DeDeo2003,Weller2003,Bean2004,Hannestad2005,Mota2007,Xia2008,Dent2009,Grande2009,Christopherson2009,Sapone2009,Putter2010,Ballesteros2010,Haq2011,Piattella2014} (see also books \cite{Amendola2010,Wolschin2010,Ruiz2010,Novosyadlyj2013m} and citing therein). They can be summarized as follows: a) the evolution of dark energy density perturbations depends on the value of effective sound speed: their amplitudes  increase when the scales of perturbations are larger than acoustic horizon scale ($k^{-1}>c_st$) and decay when they become smaller ($k^{-1}<c_st$), b) practically for any $0<c_s^2\le1$ the amplitudes of density perturbations of dark energy are essentially lower than the amplitudes for dark matter and baryon 
components at current epoch, c) the value of EoS parameter as well as the character of its time variation changes the evolution of density perturbations too: the lower initial value of $w_{de}$, the lower initial amplitude of scalar field density perturbations.

In this paper we discuss the possibility of reconstruction of Lagrangian of the scalar field dark energy with constant effective sound speed. We discuss also the possibility of determination of the effective sound speed of quintessential/phantom model of dark energy, for which in case of $c_s^2=1$ the phantom is preferred by the current observational data. For this type of dark energy the problem of determination of the effective sound speed is complicated by the too weak influence of dark energy perturbations on the matter ones. If the  phantom type of dark energy is confirmed then for its reliable reconstruction not only the increasing requirements for accuracy of data, but also radically new ideas for its study would be necessary.

In the analysis we use the minimally coupled scalar field model of dynamical dark energy that can be either quintessential or phantom with barotropic EoS \cite{Novosyadlyj2010,Sergijenko2011,Novosyadlyj2012,Novosyadlyj2013}. The existence of analytical solutions for evolution of such a scalar field, their regularity and applicability for any epoch in the past as well as in the future could make it a useful model to establish the type of dark energy.

The paper is organized as follows: in Sect. \ref{reconstr} we analyze the principal possibility of reconstruction of Lagrangian of the scalar field dark energy; in Sect. \ref{effects} we analyze the gravitational instability of the scalar field with generalized linear barotropic equation of state and constant effective sound speed and its effects on structure formation; in Sect. \ref{best-fit} we present the results of estimation of 4 dark energy parameters ($\Omega_{de}$, $w_0$, $cs^2$ and $c_s^2$). The conclusions are presented in Sect. \ref{conclusions}. 

\section{Reconstruction of Lagrangian of the scalar field dark energy with constant effective sound speed}\label{reconstr}

Scalar field dark energy can be described in several ways. Among them the most popular one is the phenomenological approach in which the dark energy is assumed to be the perfect fluid described by a small number of parameters, e. g. 3 for a constant equation of state parameter ($\Omega_{de}$, $w_{de}$ and $c_s^2$). This is convenient for practical calculations and putting the observational constraints on model parameters, however gives very little information about the physical nature of dark energy. On the other hand, the scalar field approach is well suited for study of physics of dark energy, but is not as usable in practice as the former one. So it is often useful to combine both methods of modeling of dark energy and use in numerical calculations the phenomenological perfect fluid, while study the physical features of scalar fields reconstructed to mimic the behavior of this perfect fluid.

It is well-known that the potential of a scalar field with given Lagrangian can be uniquely reconstructed using only the dark energy density and its equation of state: from their temporal dependences it is easy to determine the temporal dependences of the field potential and its kinetic term. From the temporal dependence of a kinetic term it is possible to obtain the temporal dependence of a field variable which allows the determination of explicit dependence of a potential on field variable. Note that the precision of reconstruction of the potential from observational data is limited by the precision of cosmological parameters estimation (e. g. for the discussion of how the uncertainties in determination of different parameters affect the reconstructed potential in the case of $w=const$ see \cite{Sergijenko2008}). However, the infinite degeneracy of the forms of Lagrangians exists: they lead to the same observable characteristics of dynamics of expansion of the Universe. In 2008 Unnikrishnan \cite{
Unnikrishnan2008} showed that the degeneracy is not broken even if we take into account the linear perturbations of dark energy, since different Lagrangians can lead to the same $w_{de}=const$ and $c_s^2=const$.

Let us take a closer look at the possibility of reconstruction of the functional form of Lagrangian in case of fields with $c_s^2=const$ and arbitrary (variable in time) $w_{de}$. The equation of state parameter of dark energy $w_{de}\equiv p_{de}/\rho_{de}$ and its effective (rest frame) sound speed $c_s^2\equiv\delta p_{de}^{(rf)}/\delta\rho_{de}^{(rf)}$ are defined by the field Lagrangian as follows \cite{Armendariz1999,Garriga1999}:
\begin{eqnarray}
w_{de}&=&\frac{L}{2X\frac{\partial L}{\partial X}-L},\label{w}\\
c_s^2&=&\frac{\frac{\partial L}{\partial X}}{2X\frac{\partial^2 L}{\partial X^2}+\frac{\partial L}{\partial X}}.\label{cs2}
\end{eqnarray}
Assuming that $c_s^2=const$ we obtain from (\ref{cs2}) the general form of the Lagrangian:
\begin{eqnarray}
L=VX^{\frac{1+c_s^2}{2c_s^2}}-U,
\label{L}
\end{eqnarray}
where $U=U(\phi)$ and $V=V(\phi)$ are the potentials and $X=\phi_{;i}\phi^{;i}/2$ is the kinetic term for the field $\phi$. The temporal dependence of the potential $U$ is unambiguously determined from (\ref{w}) as:
\begin{eqnarray}
U=\frac{\rho_{de}(c_s^2-w_{de})}{1+c_s^2}.
\label{U}
\end{eqnarray}
We see that the product $VX^{\frac{1+c_s^2}{2c_s^2}}$ is also determined unambiguously, but $V$ and $X$ separately are not. 
\begin{figure*}[tbp]
\includegraphics[width=0.49\textwidth]{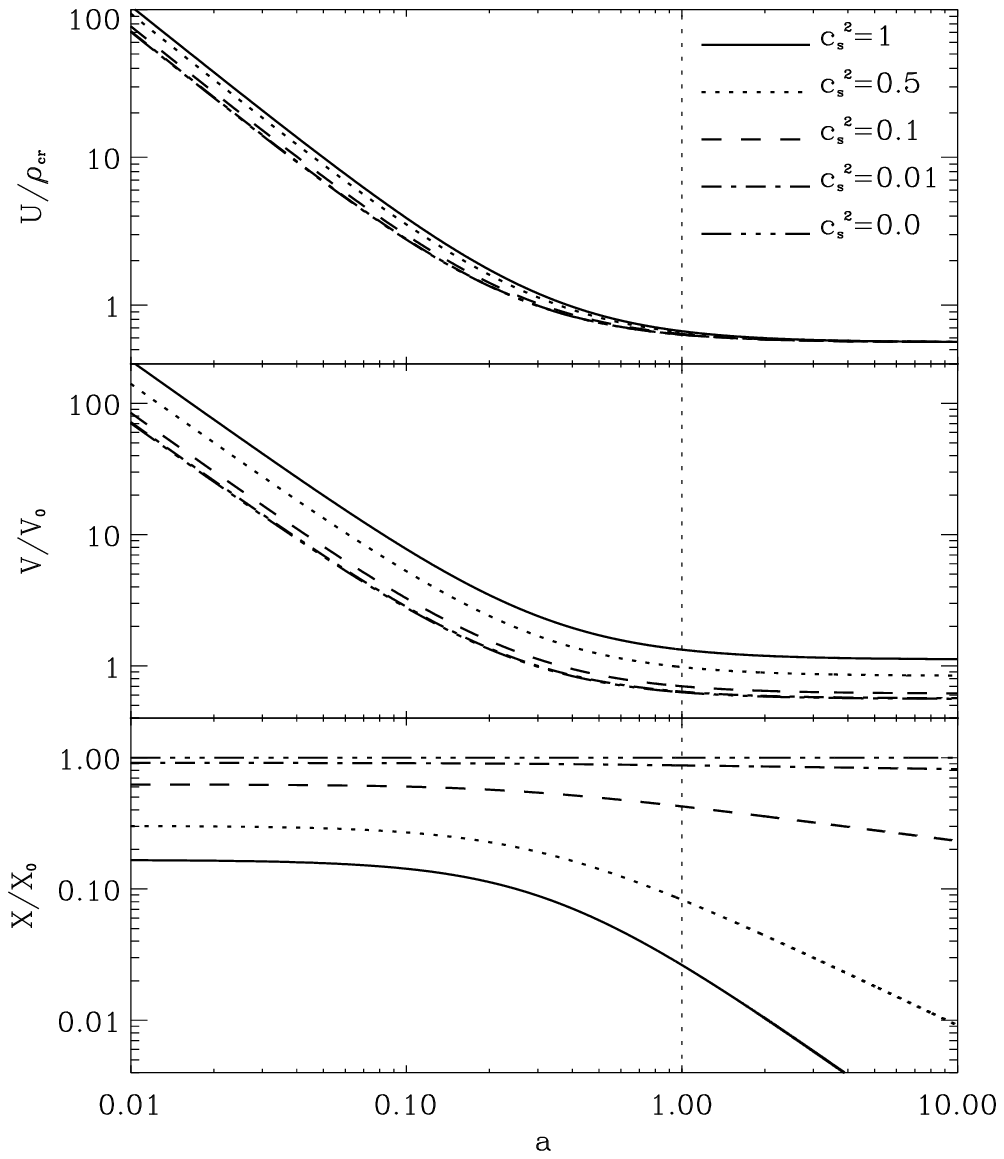}
\includegraphics[width=0.49\textwidth]{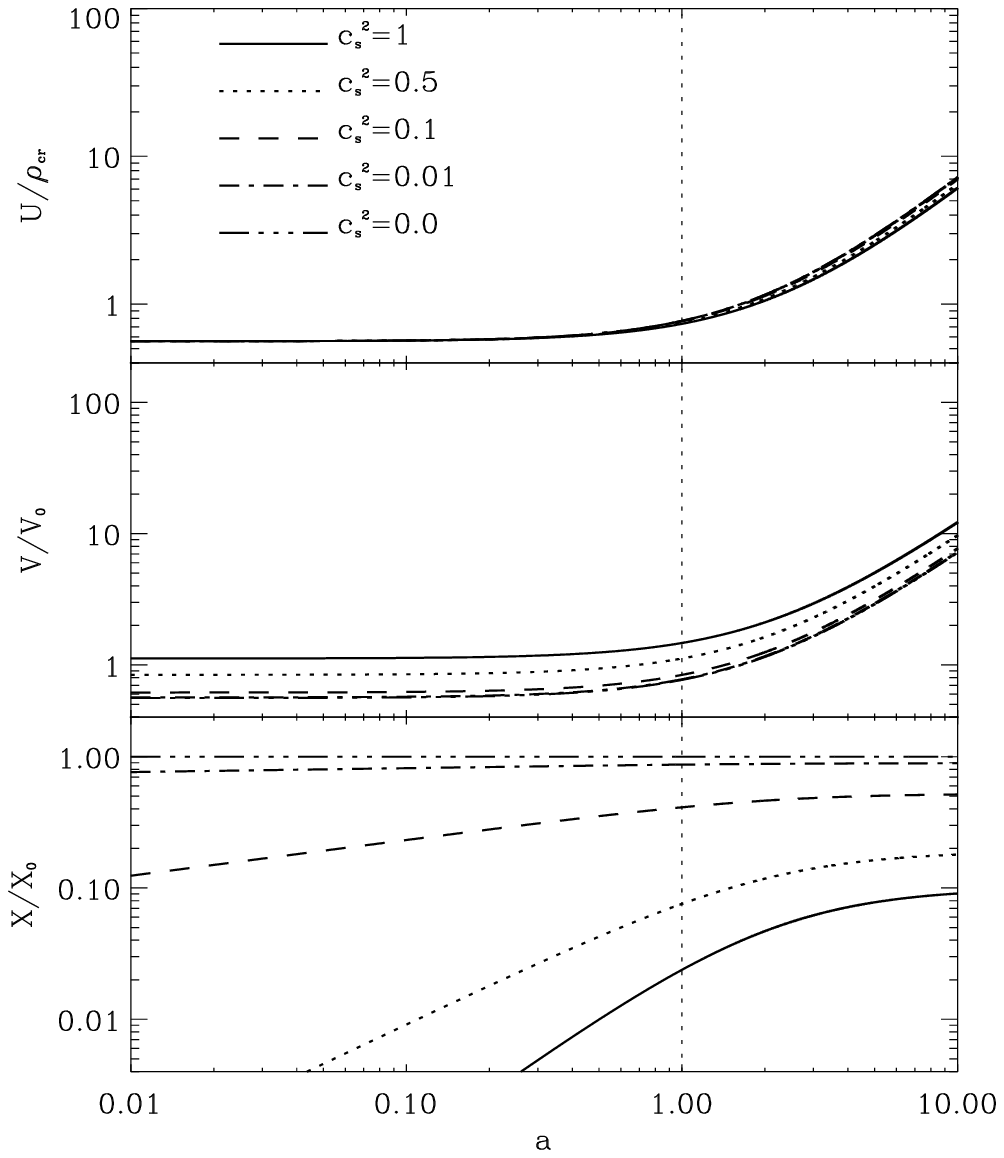}
\caption{The dependences of $U$, $V$ and $X$ on the scale factor $a$ for different values of $c_s^2$ (left panel -- quintessential scalar field with $w_0=-0.9$, $c_a^2=-0.5$, right panel -- phantom scalar field with $w_0=-1.1$, $c_a^2=-1.5$; for both $\Omega_{de}=0.7$ and $X_0\equiv(\pm V_0)^{-\frac{2c_s^2}{1+c_s^2}}$, sign ``+'' stands here for quintessence, $V_0>0$,  sign ``--'' for phantom, $V_0<0$). Vertical dotted lines mark the current epoch.}
\label{uxv}
\end{figure*}
The explicit expressions for their temporal dependences can be obtained in the form: 
\begin{eqnarray}
V&=&V_0(w_{de}-c_s^2)\rho_{de},\\
X&=&\left(\frac{1}{V_0}\frac{c_s^2}{1+c_s^2}\frac{1+w_{de}}{w_{de}-c_s^2}\right)^{\frac{2c_s^2}{1+c_s^2}}.
\label{VX}
\end{eqnarray}
Here $V_0$ is an arbitrary integration constant, for determination of which we have no condition. So, as expected, in the case of constant $c_s^2$ the infinite degeneracy of reconstructed Lagrangians cannot be broken. For the special case $c_s^2=1$ the Lagrangian (\ref{L}) can be reduced to the canonical form $L=X-U$.

The dependence of dark energy density on time or scale factor $a$ is obtained by integration of the continuity equation $T^i_{0;i}=0$ and has the general form for any dependence of EoS parameter $w_{de}$ on the scale factor: 
\begin{eqnarray}
\label{rho_de}
\rho_{de}=\rho_{de}^{(0)}a^{-3(1+\tilde{w}_{de})}, \\ 
\tilde{w}_{de}=\frac{1}{\ln{a}}\int_{1}^{a}{w_{de}(\tilde{a})d\ln{\tilde{a}}},
\end{eqnarray}
where the dark energy density at current epoch $\rho_{de}^{(0)}$ is determined by the dimensionless parameter $\Omega_{de}\equiv8\pi G \rho_{de}^{(0)}/3H_0^2$. For the constant EoS parameter $\tilde{w}_{de}=w_{de}$. In this paper we consider the scalar field model with generalized linear barotropic EoS $p_{de}=c_a^2\rho_{de}+C$ \cite{Babichev2005,Holman2004}, where $c_a^2\equiv\dot{p}_{de}/\dot{\rho}_{de}$ and $C$ are arbitrary constants defining the dynamical properties of scalar field on the cosmological background. The analytical dependences of $w_{de}$ and $\rho_{de}$ on $a$ have been obtained in \cite{Novosyadlyj2012,Novosyadlyj2013} and are as follows:
\begin{eqnarray}
w_{de}&=&\frac{(1+c_a^2)(1+w_0)}{1+w_0-(w_0-c_a^2)a^{3(1+c_a^2)}}-1,\label{rhow}\\ 
\rho_{de}&=&\rho_{de}^{(0)}\frac{(1+w_0)a^{-3(1+c_a^2)}+c_a^2-w_0}{1+c_a^2},
\label{w_de} 
\end{eqnarray}
where $w_0$ is the EoS parameter at the current epoch, $a=1$. They essentially simplify the analysis without reducing generality. For such scalar field its phenomenological density $\rho_{de}$ and pressure $p_{de}$ are analytical functions of $a$ for any values of the constants $c_a^2$ and $w_0$ ($C=\rho_{de}^{(0)}(w_0-c_a^2)$) defining its type and dynamics. Both have the clear physical meaning: $w_0$ is the EoS parameter $w_{de}$ at current epoch, $c_a^2$ is the asymptotic value of the EoS parameter $w_{de}$ at early epoch ($a\rightarrow 0$) for $c_a^2>-1$ and in far future ($a\rightarrow \infty$) for $c_a^2<-1$. The asymptotic value of $w_{de}$ in the opposite time direction is $-1$ in both cases. So, the Lagrangian of such scalar field model of dark energy can be reconstructed accurately up to a constant $V_0$ if parameters $\Omega_{de}$, $w_0$, $c_a^2$ and $c_s^2$ are given or determined using observational data. In Fig. \ref{uxv} we present the dependences $U(a)$, $V(a)$ and $X(a)$ for different values 
of $c_s^2$. 

One can see that the potential $U(a)$ changes slightly with the value of $c_s^2$ from the range [0,1]. Moreover, for quintessential scalar field the differences occur in the past, while for phantom scalar field in the future. The potential $V(a)$ seems to be more sensitive to $c_s^2$, but indefiniteness of constant $V_0$ cancels this advantage: one can renormalize the potentials so that these lines superimpose. The situation is better for kinetic terms $X(a)$ (bottom panels): the curves for different values of effective sound speed are distinguishable in the past for both fields. So, we would hope to find some observational data which give the possibility to constrain the value of $c_s^2$ for quintessential or phantom scalar field.

\section{Effects of the effective sound speed of scalar field on CMB and large scale structure}\label{effects}

\begin{figure*}[tbp]
\includegraphics[width=0.49\textwidth]{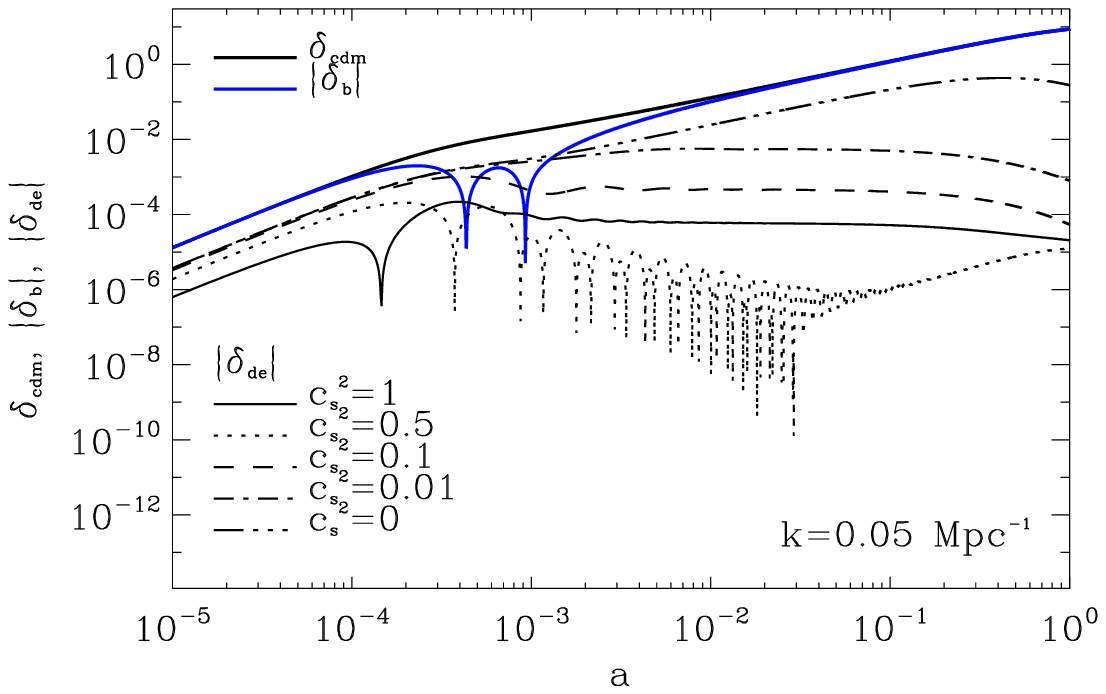}
\includegraphics[width=0.49\textwidth]{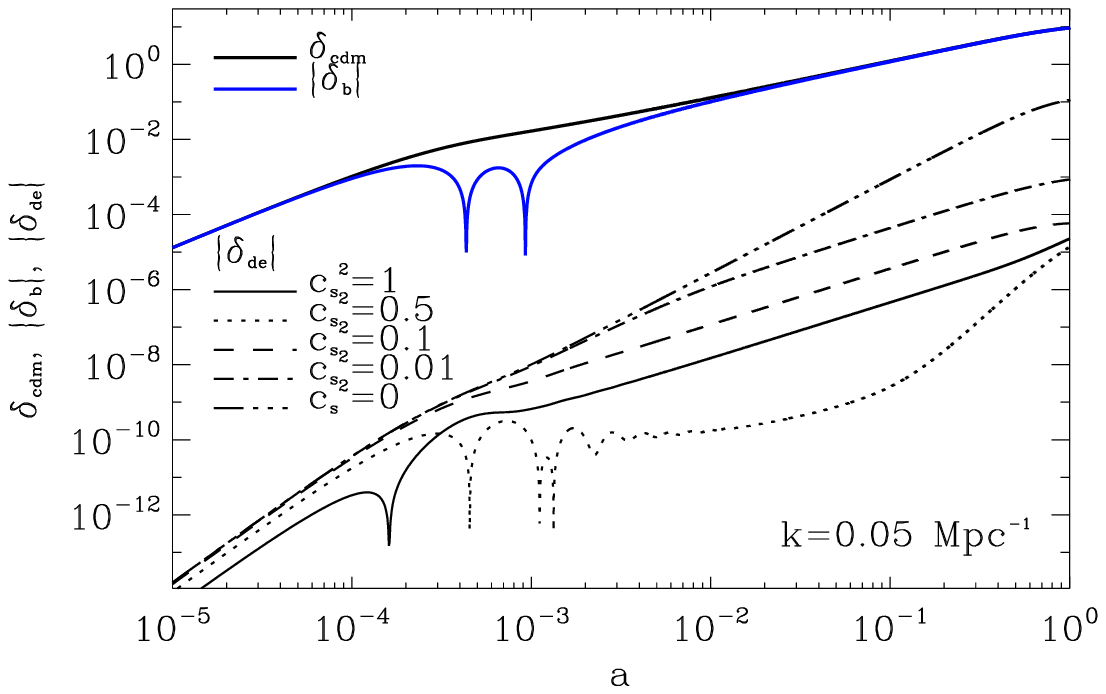}
\includegraphics[width=0.49\textwidth]{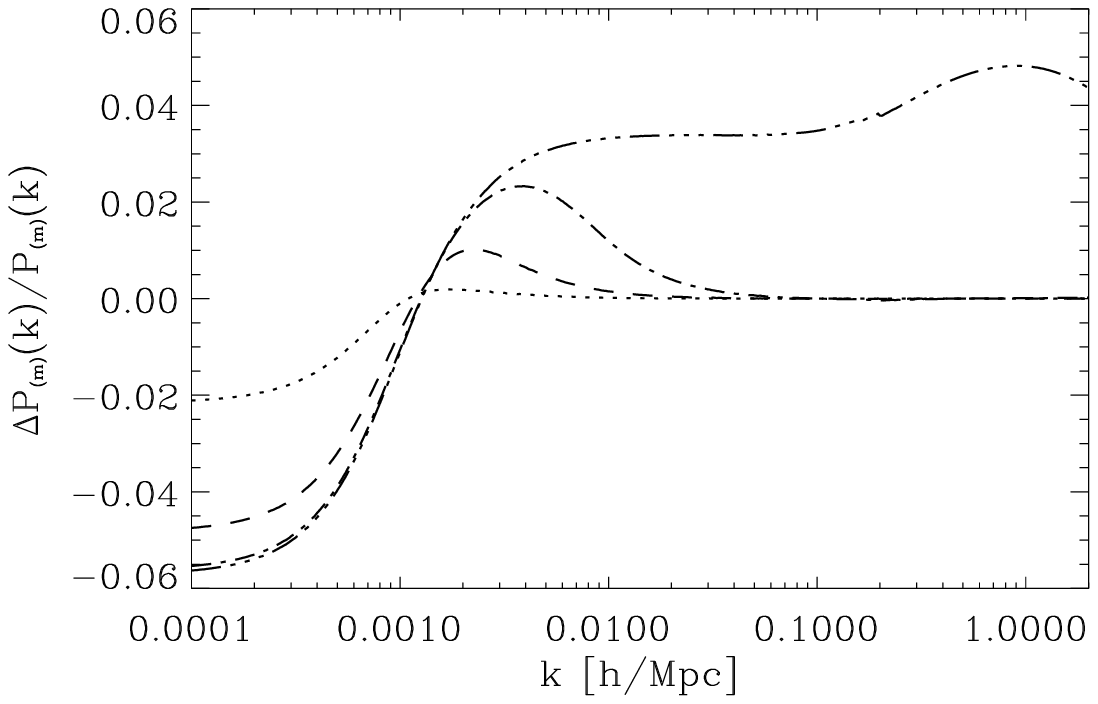}
\includegraphics[width=0.49\textwidth]{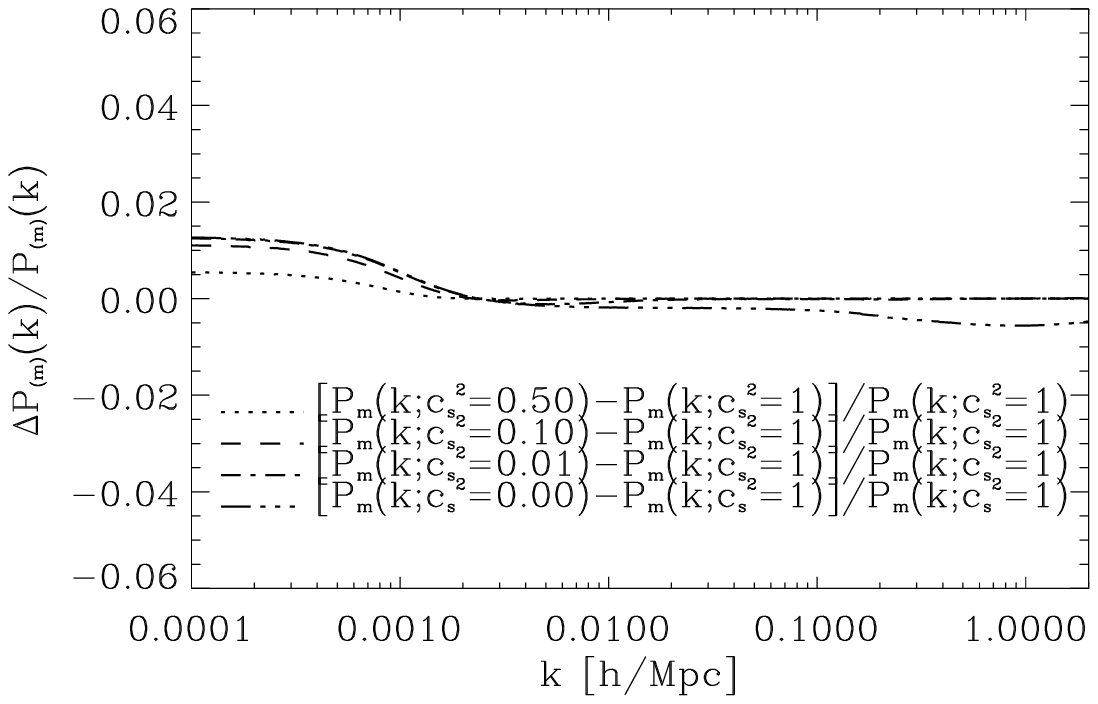}
\caption{Top panels: evolution of density perturbations of cold dark matter (top thick solid lines), baryons (blue solid lines) and scalar field dark energy with different values of effective sound speed ($c_s^2$=1, 0.5, 0.1, 0.01, 0.0). Bottom panels: effect of the sound speed of scalar field dark energy on the matter power spectrum. Left column -- quintessential scalar field with $w_0=-0.9$, $c_a^2=-0.5$; right column -- phantom scalar field with $w_0=-1.1$, $c_a^2=-1.5$. The rest of parameters correspond to the model $\mathbf{p_3}$ from \cite{Novosyadlyj2014}.}
\label{ddeb_dpk}
\end{figure*}

The value of effective sound speed of dynamical dark energy defines the evolution of density and velocity perturbations when the rest of parameters are fixed \cite{Hu1998,Erickson2002,DeDeo2003,Weller2003,Bean2004,Hannestad2005,Mota2007,Xia2008,Dent2009,Grande2009,Christopherson2009,Sapone2009,Putter2010,Ballesteros2010,Haq2011,Piattella2014,Amendola2010,Wolschin2010,Ruiz2010,Novosyadlyj2013m}. For illustration of this effect we integrate the evolution equations for perturbations of densities and velocities of each component (dark energy, cold dark matter, baryons, thermal electromagnetic radiation and active neutrinos) and metric in synchronous gauge. The evolution of perturbations for the scalar field with (\ref{rhow})-(\ref{w_de}) and any $c_s^2$ can be described by the system of differential equations:
\begin{eqnarray}
\delta'_{de}+3\frac{c_s^2-w_{de}}{a}\delta_{de}+\frac{1+w_{de}}{2}h'+\nonumber&&\\
(1+w_{de})\left[\frac{k}{a^2H}+9H\frac{c_s^2-c_a^2}{k}\right]v_{de}=0,&&\label{d_de-perturb}\\
v'_{de}+\frac{1-3c_s^2}{a}v_{de}-\frac{c_s^2 k}{(1+w_{de})a^2H}\delta_{de}=0,&&\label{v_de-perturb}\\
h''+\frac{2-q}{a}h'=-\frac{3}{a^2}\sum_{i}\Omega_i\left[(1+3c_{s(i)}^2)\delta_i+\right.\nonumber&&\\
\left.9aH(1+w_i)(c_{s(i)}^2-c_{a(i)}^2)\frac{v_i}{k}\right],&&\label{h-perturb}
\end{eqnarray}
where $\delta_{i}$ and $v_{i}$ are the Fourier amplitudes of density and velocity perturbations for $i$-component, $h\equiv h^j_j$ is the Fourier amplitude of metric perturbations. Here $(')\equiv d/da$, $H\equiv H_0\sqrt{\sum_{i}\Omega_i^{(0)}a^{-3(1+\tilde{w}_i)}}$, $q\equiv \frac{1}{2}\sum_{i}\Omega_i(1+3w_i)$, $\Omega_i\equiv \rho_i(a)/\sum_{i}{\rho_i(a)}$ and for components with $w=const$ $\tilde{w}=w$. The equations for cold dark matter can be obtained from (\ref{d_de-perturb})-(\ref{v_de-perturb}) assuming $w=c_a=c_s=0$. The equations for density and velocity perturbations of baryons, thermal electromagnetic radiation and neutrinos and are presented in \cite{Ma1995}. Equations (\ref{d_de-perturb})-(\ref{v_de-perturb}) show that the effective sound speed influences the evolution of density and velocity perturbations of dark energy.  

Taking the adiabatic initial conditions for matter perturbations and treating the dark energy as test component in the gravitational potential of matter we obtain the relations between Fourier amplitudes at some $a_{init}\ll1$: 
\begin{eqnarray}
 \delta_{de}^{(init)}&=&-\frac{(1+w_{de})(4-3c_s^2)h^{(init)}}{8+6c_s^2-12w_{de}+9c_s^2(w_{de}-c_a^2)},\\
 v_{de}^{(init)}&=&-\frac{c_s^2k\eta_{init}h^{(init)}}{8+6c_s^2-12w_{de}+9c_s^2(w_{de}-c_a^2)},
\end{eqnarray}
where $\eta_{init}$ is the conformal time at $a_{init}$. So, in the early Universe, when the scale of perturbation is superhorizon, the matter density perturbations and the quintessential scalar field ($1+w_{de}>0$) ones have the same sign: positive matter density perturbation -- positive scalar field one, negative matter density perturbation -- negative scalar field one. The amplitude of dark matter density perturbations increases monotonically changing the rate at the transition from radiation-dominated epoch to matter-dominated one. The baryon-photon density perturbation starts to oscillate after entering into sound horizon and continues up to the recombination epoch. The temporal behavior of scalar field density perturbations is more complicated since it depends on the value of effective sound speed of scalar field and the wave number of perturbation. We illustrate this by calculations.

In the top panels of Fig. \ref{ddeb_dpk} we present the results of integration of such equations by CAMB \cite{camb,camb_source} in cosmological model with the quintessential scalar field (left) and the phantom one (right) for different values of effective sound speed: $c_s^2$=1, 0.5, 0.1, 0.01, 0.0. All perturbations are computed in the cold dark matter rest frame. The density perturbations of cold dark matter ($\delta_{cdm}$) and baryons ($\delta_{b}$) are presented only for the model with dark energy with $c_s^2$=1 (top thick solid lines). For the scalar field the absolute values of density perturbations ($|\delta_{de}|$) are presented. As it is shown for $k=0.05$ Mpc$^{-1}$, the amplitude of $\delta_{de}$ changes the sign from ``+'' to ``-'' after entering its own acoustic horizon and then freezes at some value for $c_s^2=1$, changes the sign from ``+'' to ``-'' after entering its own acoustic horizon and then freezes at some value after few oscillations for $c_s^2\sim0.5$, does not change the sign but 
freezes at some value for $0<c_s^2<0.1$ and increases monotonically for $c_s^2=0$. 

For phantom scalar field  ($1+w_{de}<0$) the initial amplitude of $\delta_{de}$ is lower and has the sign opposite to the sign of $\delta_{m}(a_{init})$, but dependences of $|\delta_{de}(a)|$ are similar to the corresponding dependences for quintessence.

Note that for the dark energy perturbations their signs and magnitudes are strongly gauge-dependent at superhorizon scales \cite{Bean2004}.

To analyze the effect of dark energy perturbations on the matter ones we have calculated the matter power spectra (with the non-linear corrections by halofit adopted for the studied type of dark energy) for models with the same main parameters but different $c_s^2$. In the bottom panels of Fig. \ref{ddeb_dpk} we present the relative differences $P_m(k;c_s^2\ne1)/P_m(k;c_s^2=1)-1$ for $c_s^2$=0.5, 0.1, 0.01, 0.0, which illustrate the influence of reducing of the value of effective sound speed on the matter power spectrum: suppression of power at large scales ($k<0.001$ Mpc$^{-1}$) for the quintessential scalar field and enhancement for the phantom one, enhancement of power at intermediate scales ($0.001<k<0.1$ Mpc$^{-1}$) for the quintessential scalar field and small suppression for the phantom one. At $k>0.1$ Mpc$^{-1}$ the effect is completely absent for $c_s\ne0$. Therefore, the form of the matter power spectrum at large and intermediate scales is sensitive to the value of $c_s$, more sensitive for the 
quintessential scalar field and less sensitive for the phantom one. At intermediate scales, for 
which we have the observational data, the differences for models with $0\le c_s\le1$ are within 2\% for the quintessential scalar field and within 0.5\% for the phantom one, while observational uncertainties are not smaller than 10\% \cite{sdssdr7,WiggleZ} (see also Fig. 5 in \cite{Novosyadlyj2013}).

\begin{figure*}[tbp]
\includegraphics[width=0.49\textwidth]{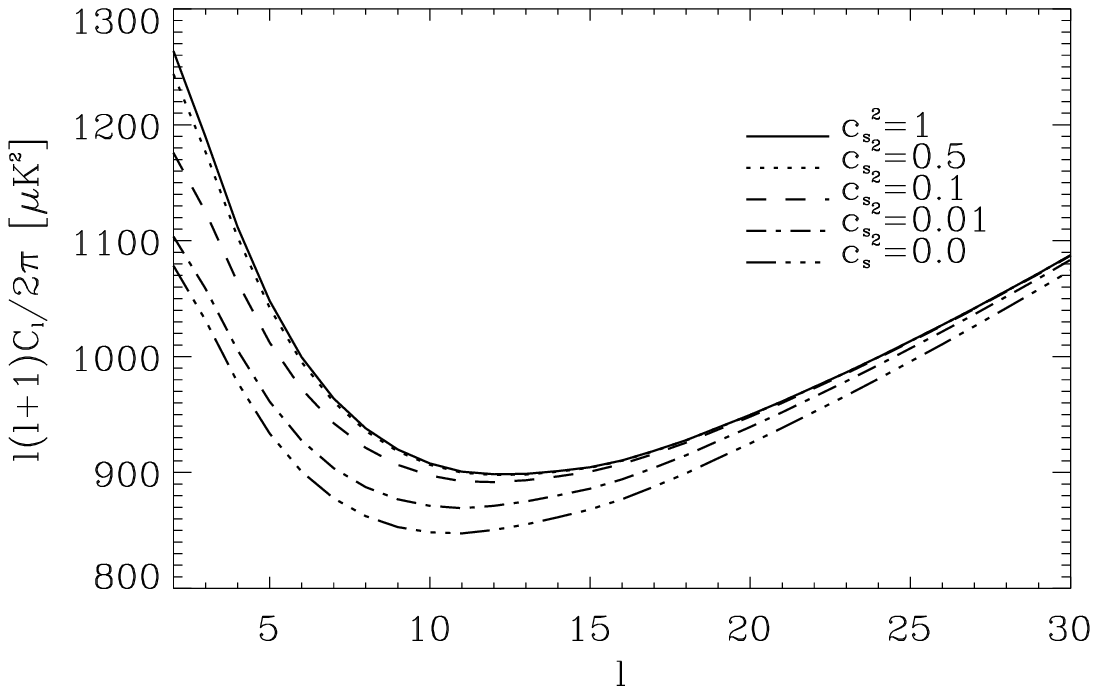} 
\includegraphics[width=0.49\textwidth]{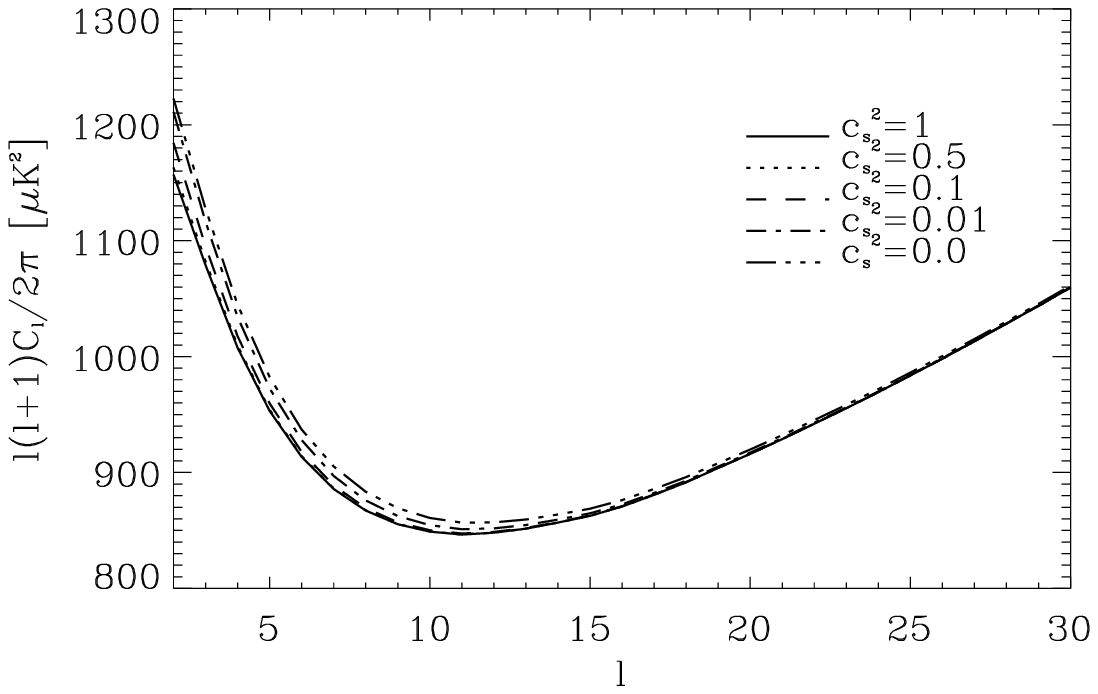}
\includegraphics[width=0.49\textwidth]{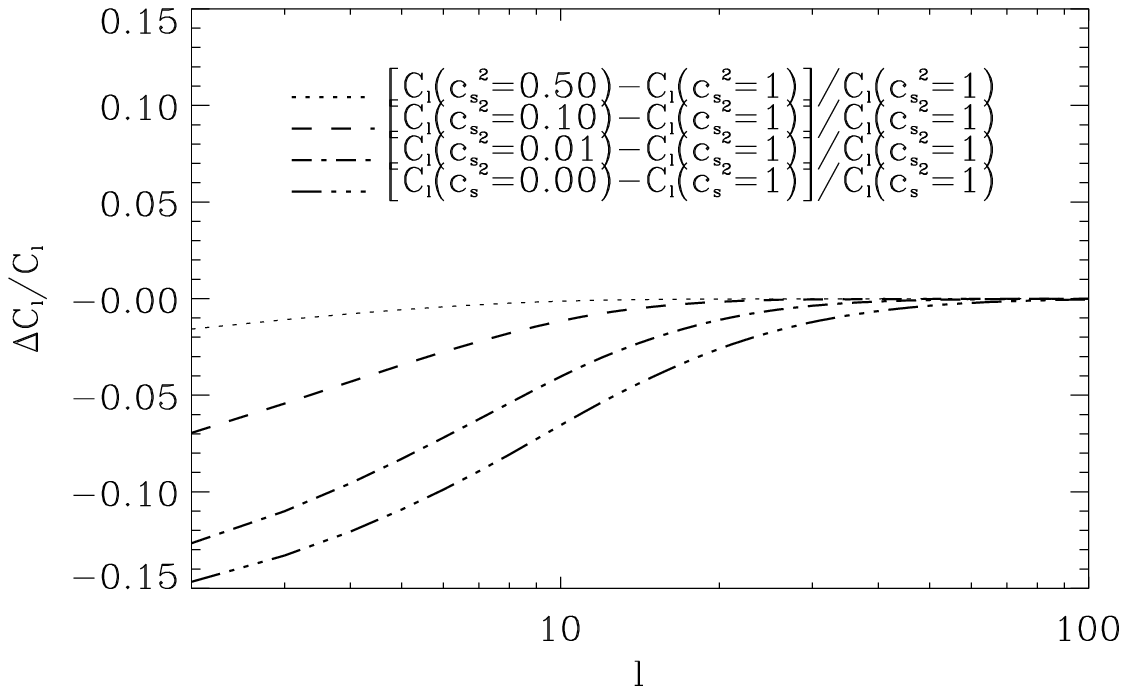}
\includegraphics[width=0.49\textwidth]{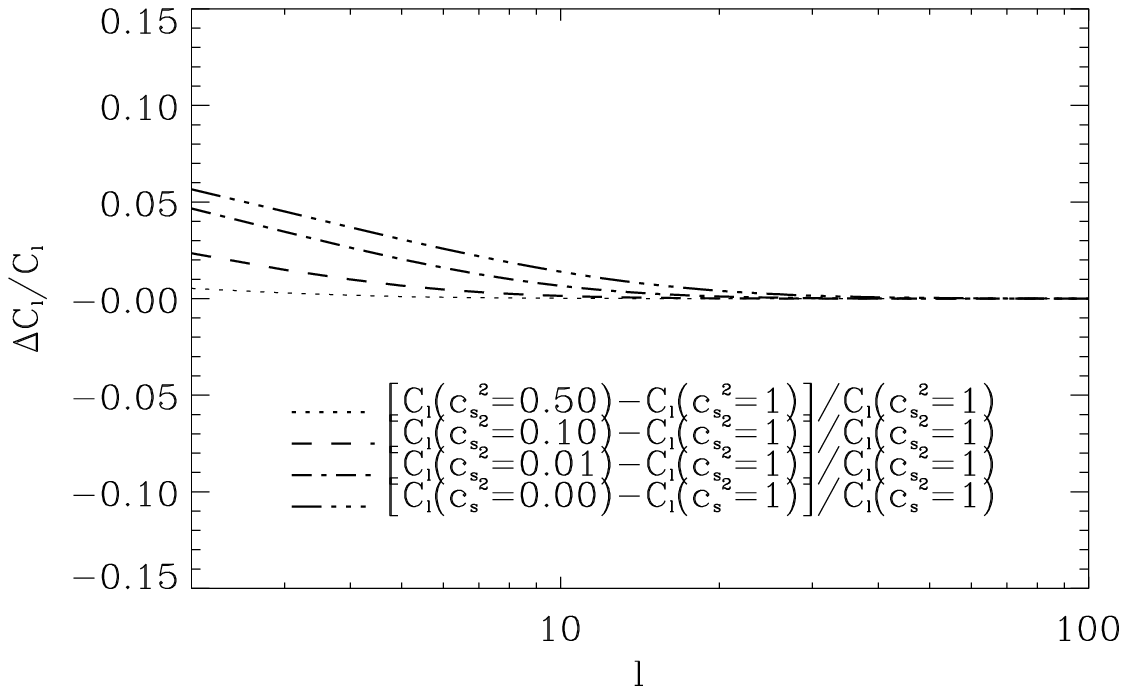}
\caption{Top panels: the angular power spectrum of CMB temperature fluctuations at large angular scales for the model with scalar field dark energy with different values of the squared effective sound speed  ($c_s^2$=1, 0.5, 0.1, 0.01, 0.0). Bottom panels: the relative differences of $C_l$'s for models with different values of the squared effective sound speed. Left column -- quintessential scalar field with $w_0=-0.9$, $c_a^2=-0.5$; right column -- phantom scalar field with $w_0=-1.1$, $c_a^2=-1.5$. The rest of parameters correspond to the model $\mathbf{p_3}$ from \cite{Novosyadlyj2014}.}
\label{cl_cs2}
\end{figure*}

\begin{figure*}[tbp]
\includegraphics[width=0.49\textwidth]{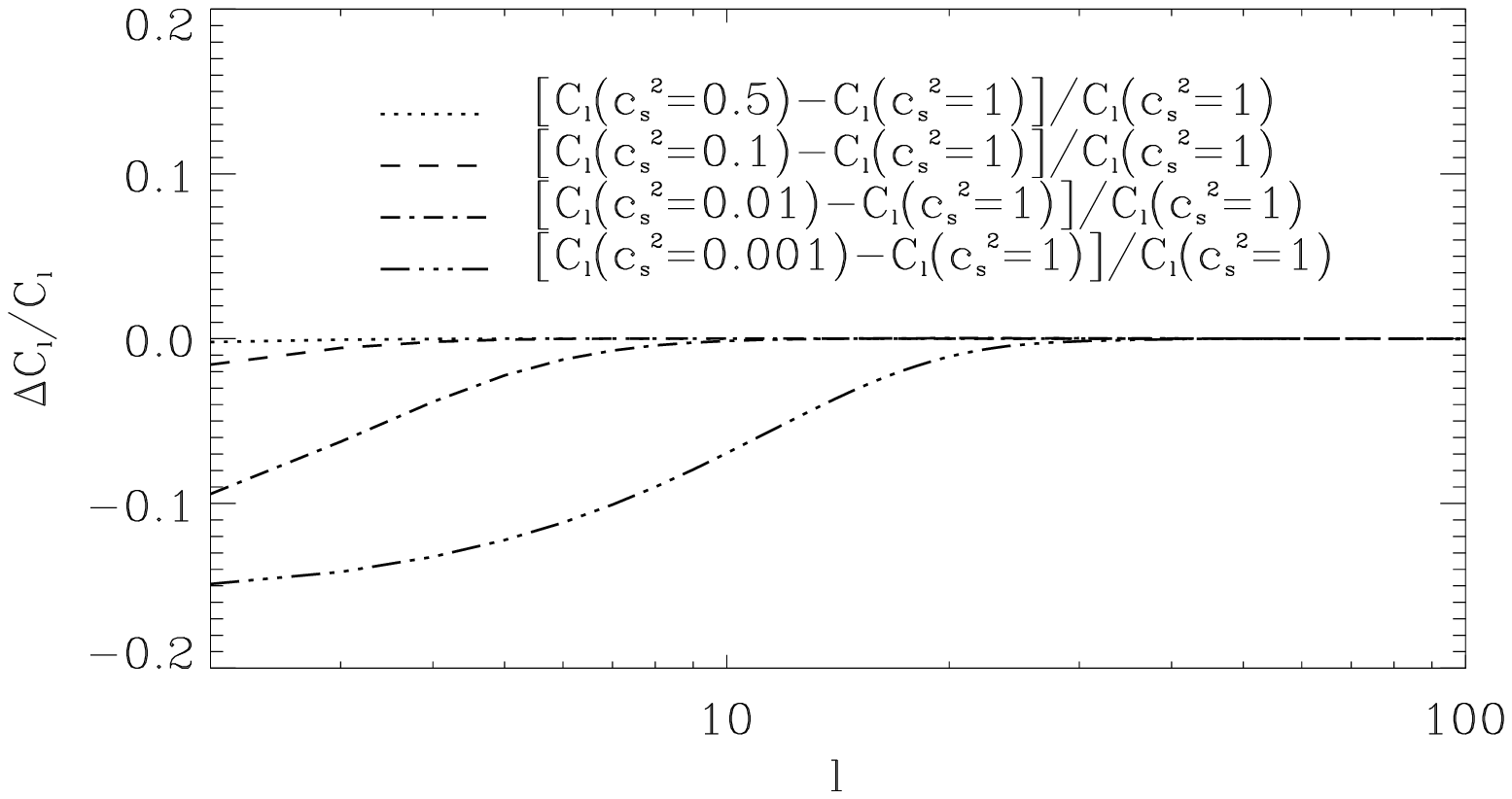} 
\includegraphics[width=0.49\textwidth]{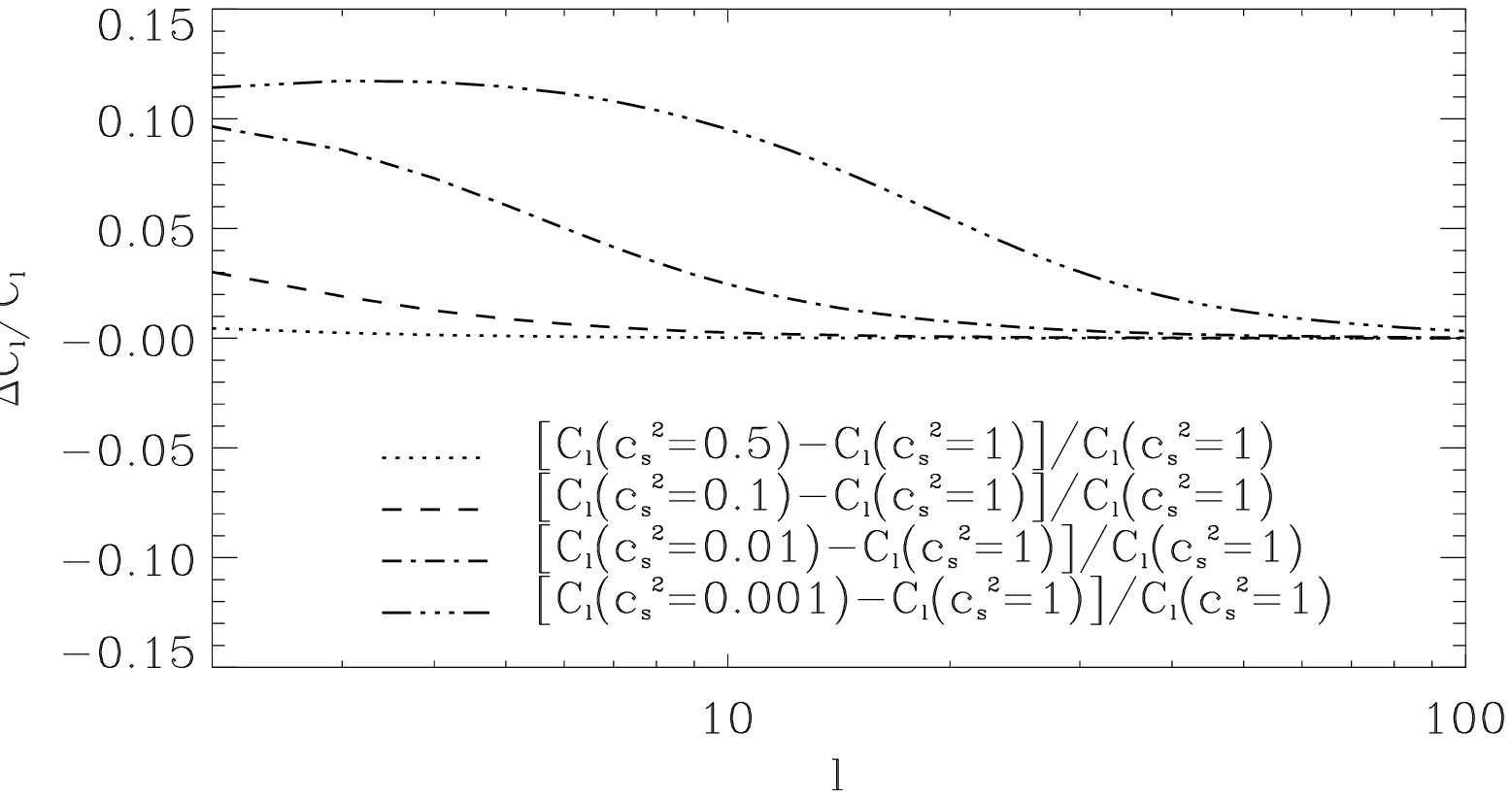}
\includegraphics[width=0.49\textwidth]{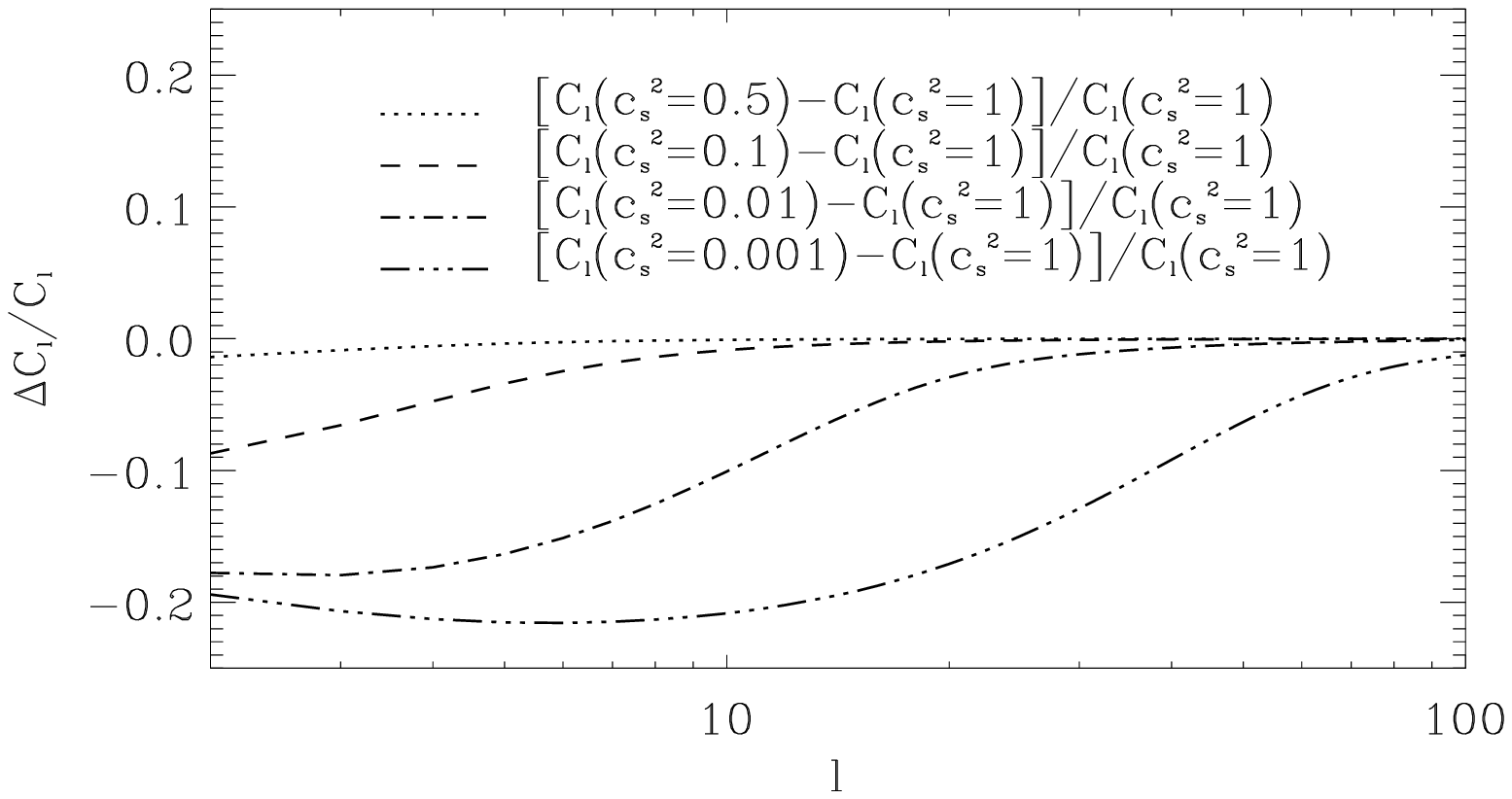}
\includegraphics[width=0.49\textwidth]{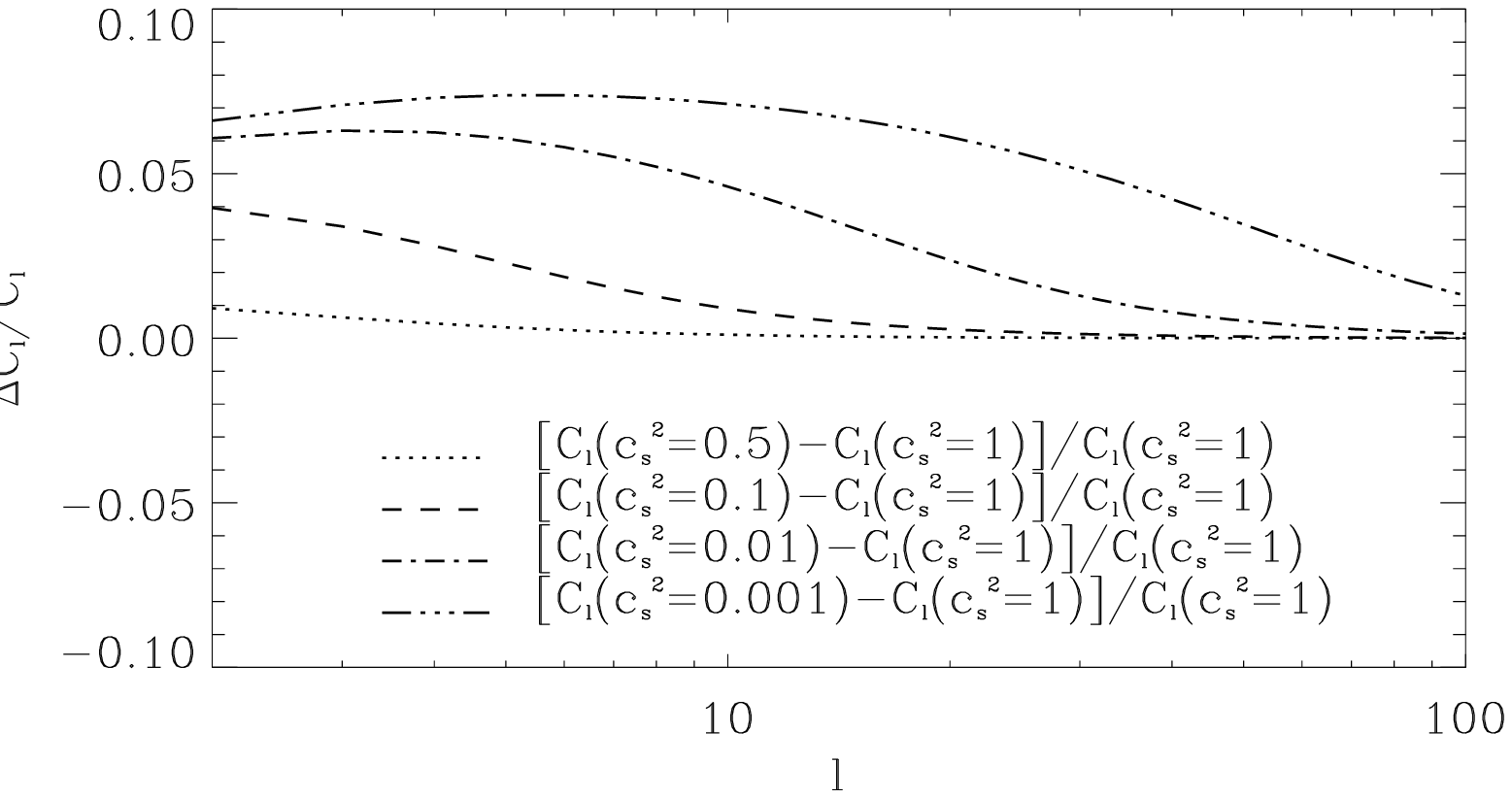}
\includegraphics[width=0.49\textwidth]{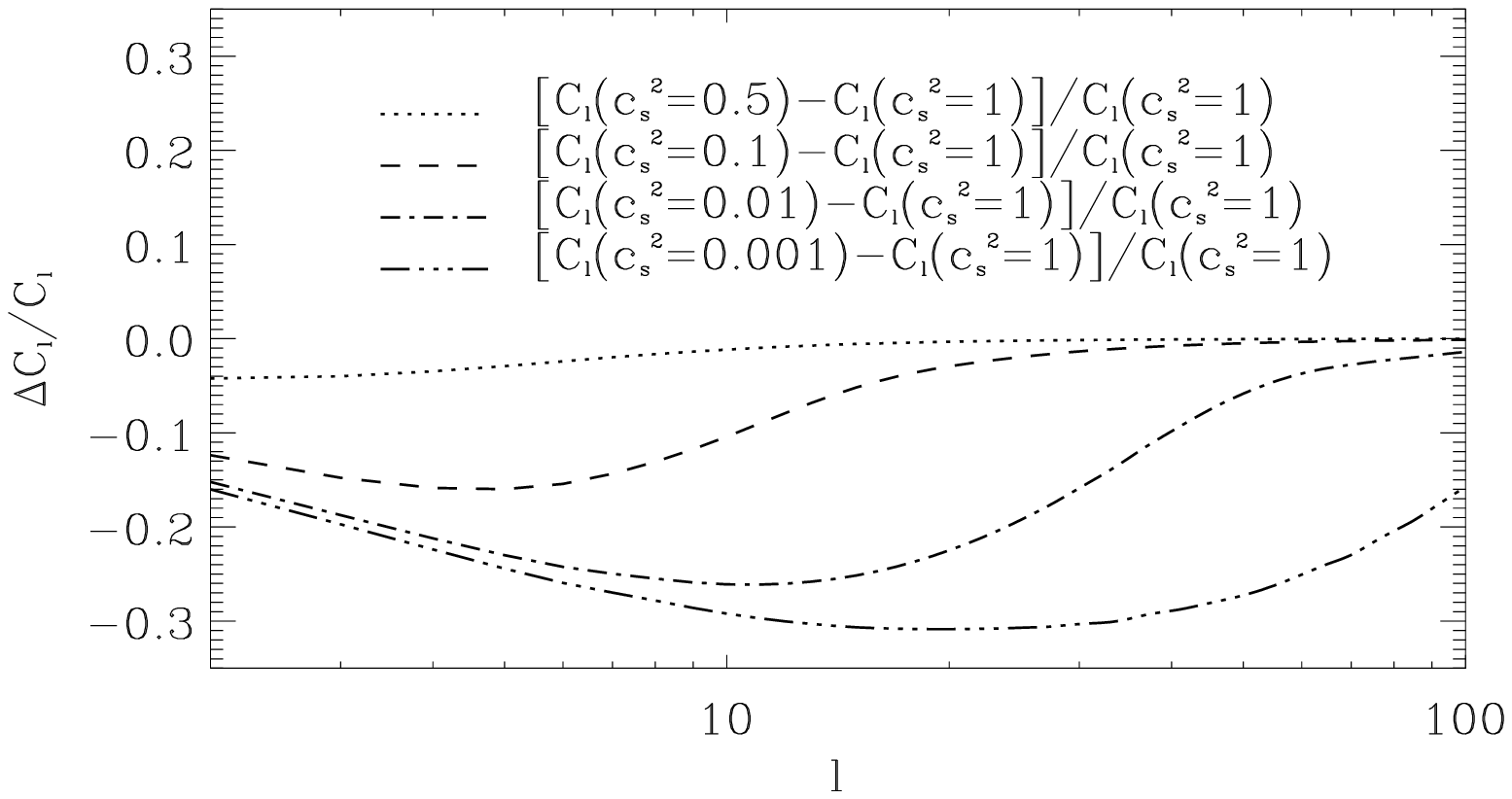}
\includegraphics[width=0.49\textwidth]{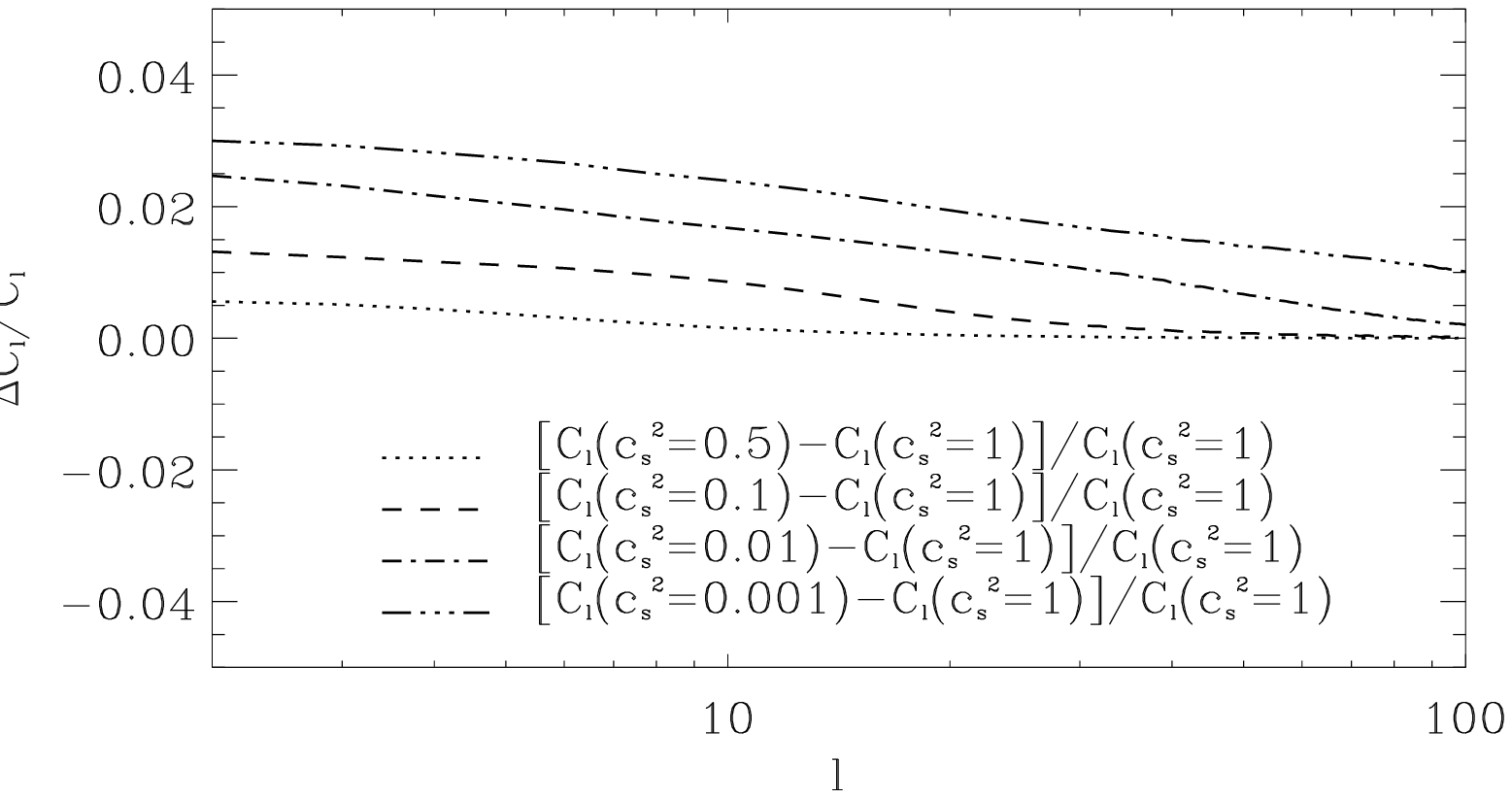}
\caption{The relative differences of angular power spectra of the CMB temperature-source number counts cross-correlation for models with different values of the squared effective sound speed. Left column -- quintessential scalar field with $w_0=-0.9$, $c_a^2=-0.5$; right column -- phantom scalar field with $w_0=-1.1$, $c_a^2=-1.5$. The rest of parameters correspond to the model $\mathbf{p_3}$ from \cite{Novosyadlyj2014}. Upper row -- Gaussian window function $W(z)$ at $z=0.3$ with $\sigma_z=0.06$ and $b=1.5$, middle -- Gaussian window function $W(z)$ at $z=1$ with $\sigma_z=0.3$ and $b=2$, bottom -- Gaussian window function $W(z)$ at $z=3$ with $\sigma_z=0.3$ and $b=2$; for all cases $s=0.42$ (for explanation of meaning of these parameters see \cite{counts2011}).}
\label{cl_cross}
\end{figure*}

The density perturbations of scalar fields with different $c_s$ should affect the temperature fluctuations of cosmic microwave background (CMB) because of Sachs-Wolfe effect \cite{SW1967} with different strength. Fig. \ref{cl_cs2} illustrates the suppression of power at large angular scales (low spherical harmonics) for the quintessential scalar field and the enhancement for the phantom one. The differences for models with $0\le c_s\le1$, shown in the bottom panels, are within 15\% for $w_0=-0.9$ and within 6\% for $w_0=-1.1$, while observational uncertainties are within 30\% \cite{wmap9a,Planck2013b} (see also Fig. 6 in \cite{Novosyadlyj2013} and Fig. 5 in \cite{Novosyadlyj2014}). So, the probing power of CMB data for $c_s$ is also weak. 

One can see that the variation of value of the effective sound speed affects the CMB temperature power spectrum at the same angular scales as the tensor mode. The uncertainty of upper limit on $r=T/S$ grows with increasing of number of degrees of freedom of the model. There was a possibility that this might reduce the tension between upper limit $r<0.11$ at 95\% C.L. given by Planck \cite{Planck2013c} and 1$\sigma$-range $0.15\le r\le0.27$ given by BICEP2 \cite{BICEP2_2014} if instead of $\Lambda$CDM the dynamical dark energy with free $c_s$ plus CDM model was used. After a series of publications \cite{Mortonson2014,Flauger2014,Bonaldi2014,planckxxx,cortes2014,cheng2014} the problem of mentioned tension was finally solved by the joint analysis of data from BICEP2/Keck Array and Planck\cite{planckbicep} giving $r_{0.05}<0.12$ at 95\% C.L. However, these speculations are interesting in view of the expected data on BB-mode polarization from Planck and other experiments.

The CMB-LSS cross-correlation data are believed to be crucial for constraining the effective sound speed of dark energy (see e. g. \cite{Ballesteros2010} who pointed out that it would not be possible to put the lower limit on $c_s^2$ or determine its order of magnitude until the availability of cross-correlation of the Planck data on CMB with the expected large scale structure data from LSST). In Fig. \ref{cl_cross} we present the relative differences for the power spectra of CMB temperature-source number counts cross-correlation (such spectra are dominated by the integrated Sachs-Wolfe effect). For detailed explanation of the method of calculation see \cite{counts2011} whose code CAMB sources \cite{camb_s_source} we have used. The suppression of power caused by the quintessence increases with redshift ($\lesssim15\%$ at $z=0.3$, $\lesssim22\%$ at $z=1$ and $\lesssim31\%$ at $z=3$) while the enhancement of power caused by the phantom decreases ($\lesssim12\%$ at $z=0.3$, $\lesssim8\%$ at $z=1$ and 
$\lesssim3\%$ at $z=3$). The relative errors of CMB-LSS cross-correlation data should be at least within 30\% to distinguish between quintessential models with different $c_s^2$ and within 10\% to distinguish between phantom ones.

Other data related to the large scales (weak gravitational lensing or cosmic shear, cosmic magnification, CMB polarization etc.) expected in the observational programs of current decade can possibly clarify the prospects of determination of the effective sound speed of dark energy (for forecasts see e. g. \cite{sapone2010}). Another possibility for obtaining the lower limit on $c_s^2$ could come from the study of dark energy distribution near the compact objects (stars, black holes) \cite{tsizh2014}.

\begin{figure*}[tbp]
\includegraphics[width=0.49\textwidth]{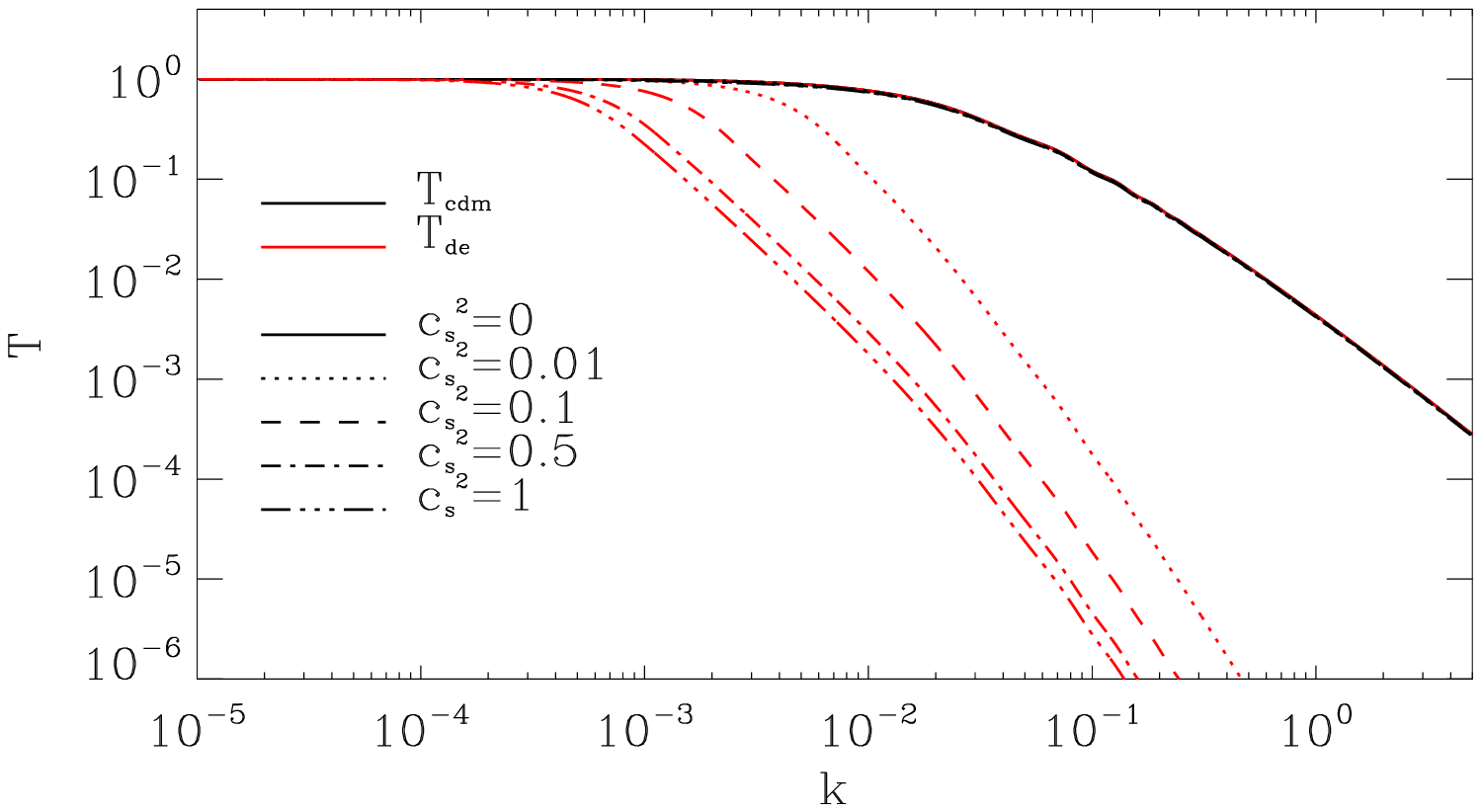} 
\includegraphics[width=0.49\textwidth]{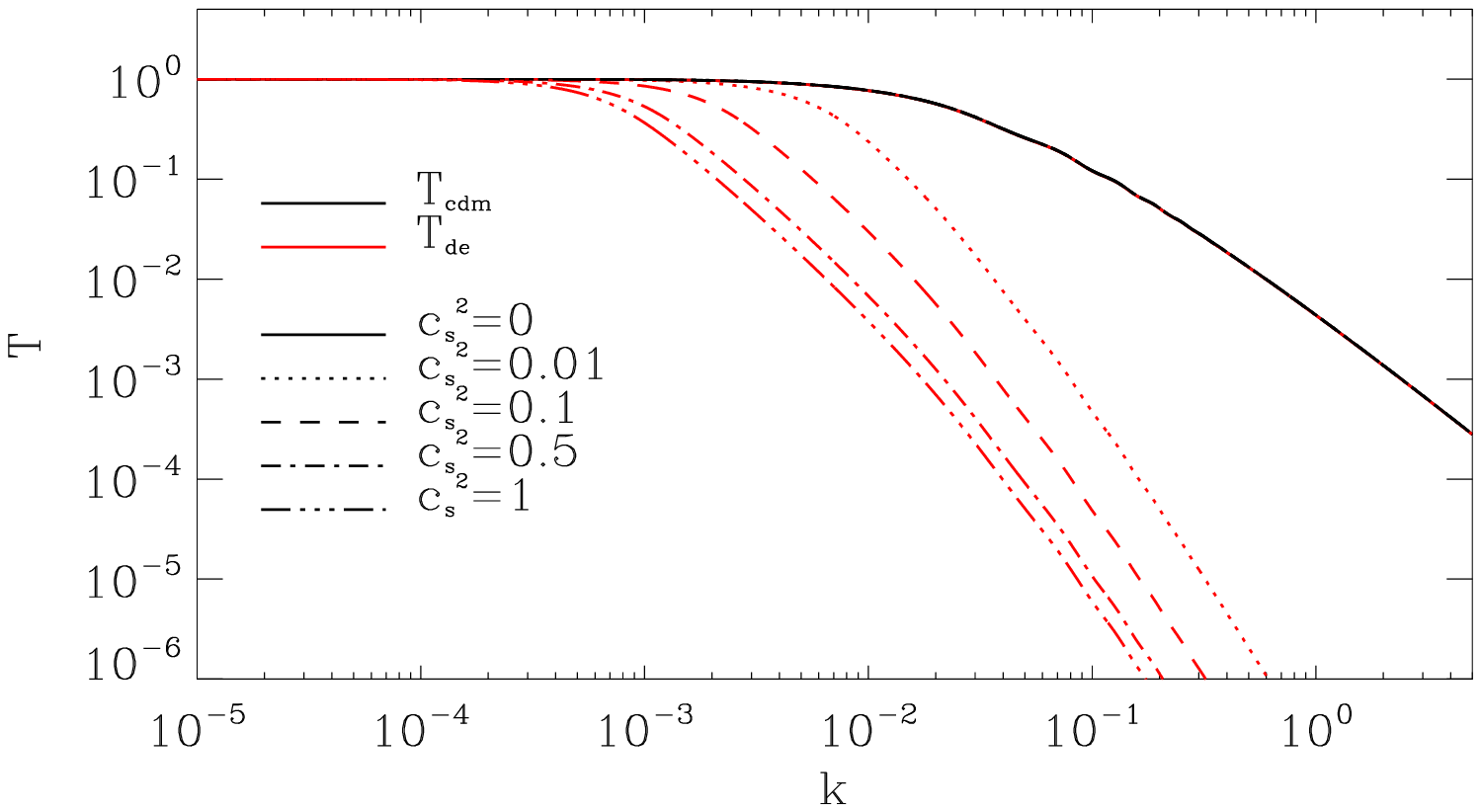}
\includegraphics[width=0.49\textwidth]{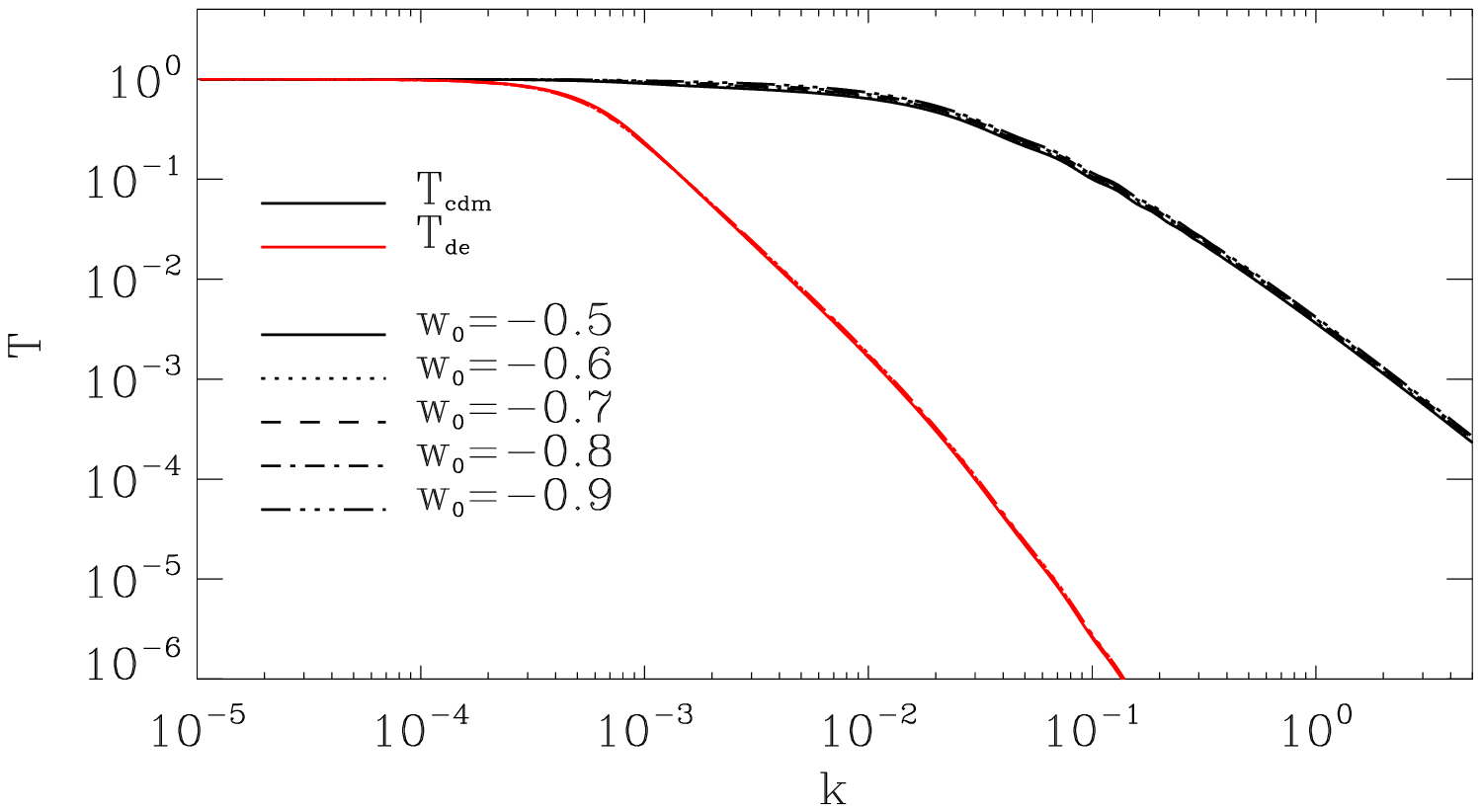}
\includegraphics[width=0.49\textwidth]{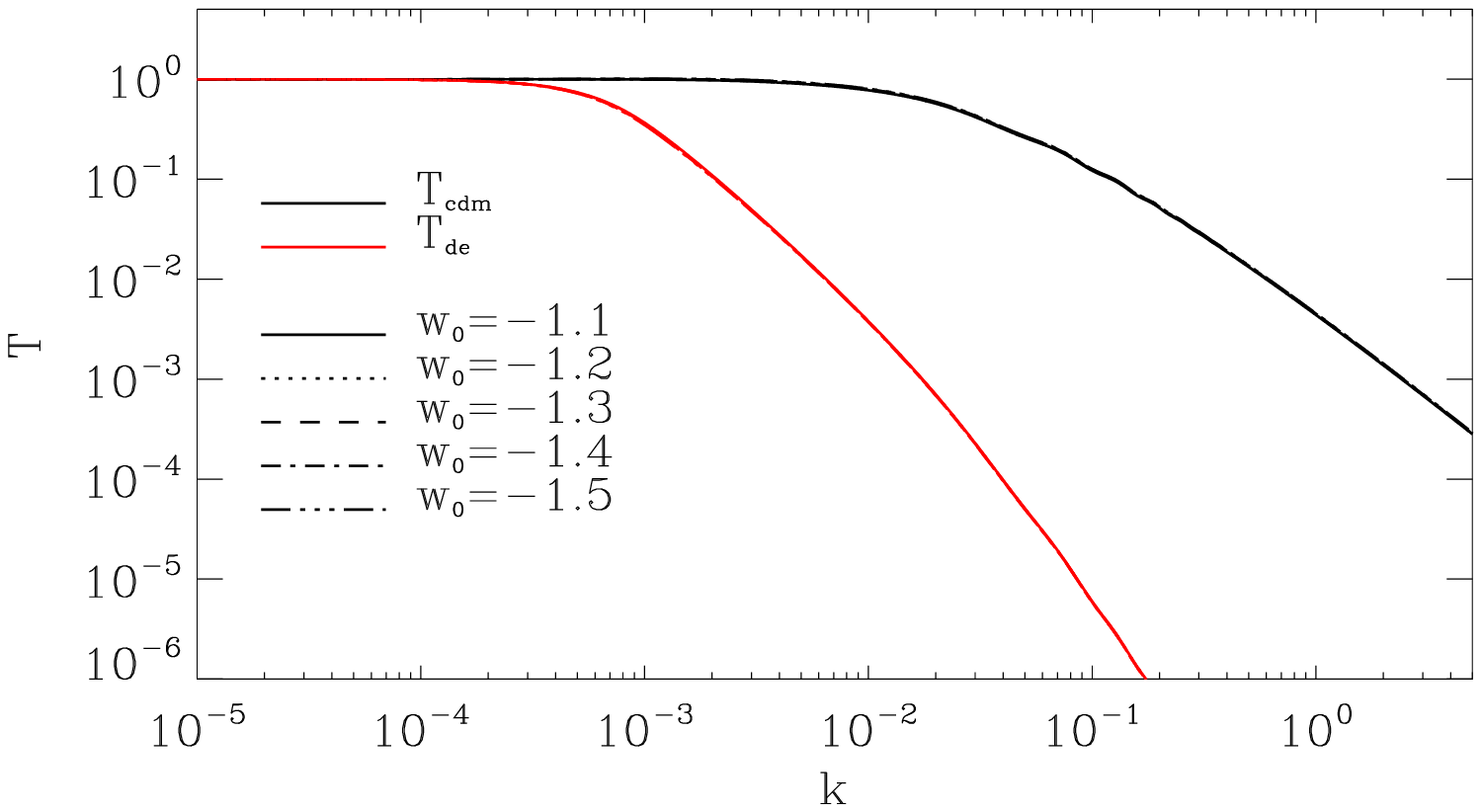}
\includegraphics[width=0.49\textwidth]{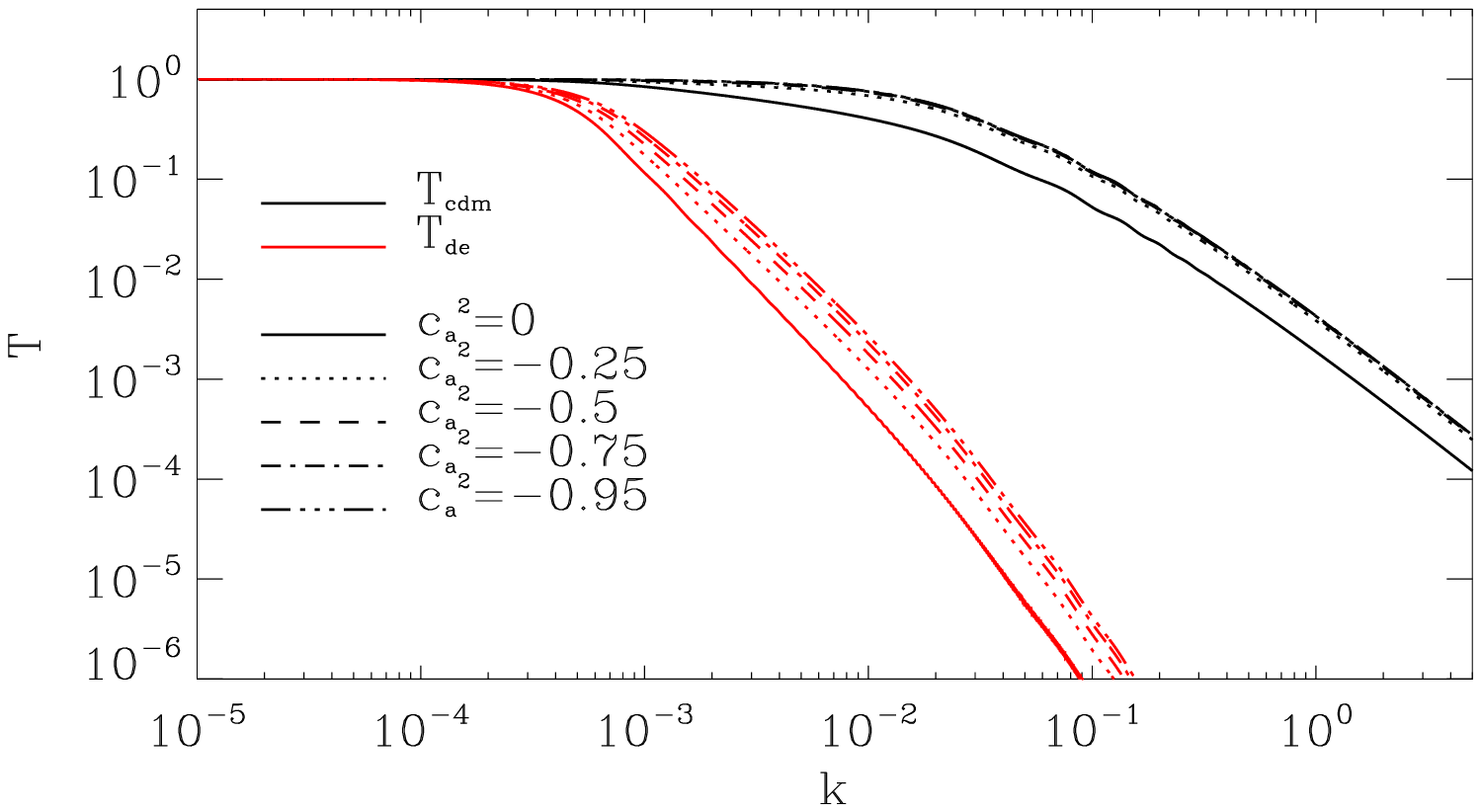}
\includegraphics[width=0.49\textwidth]{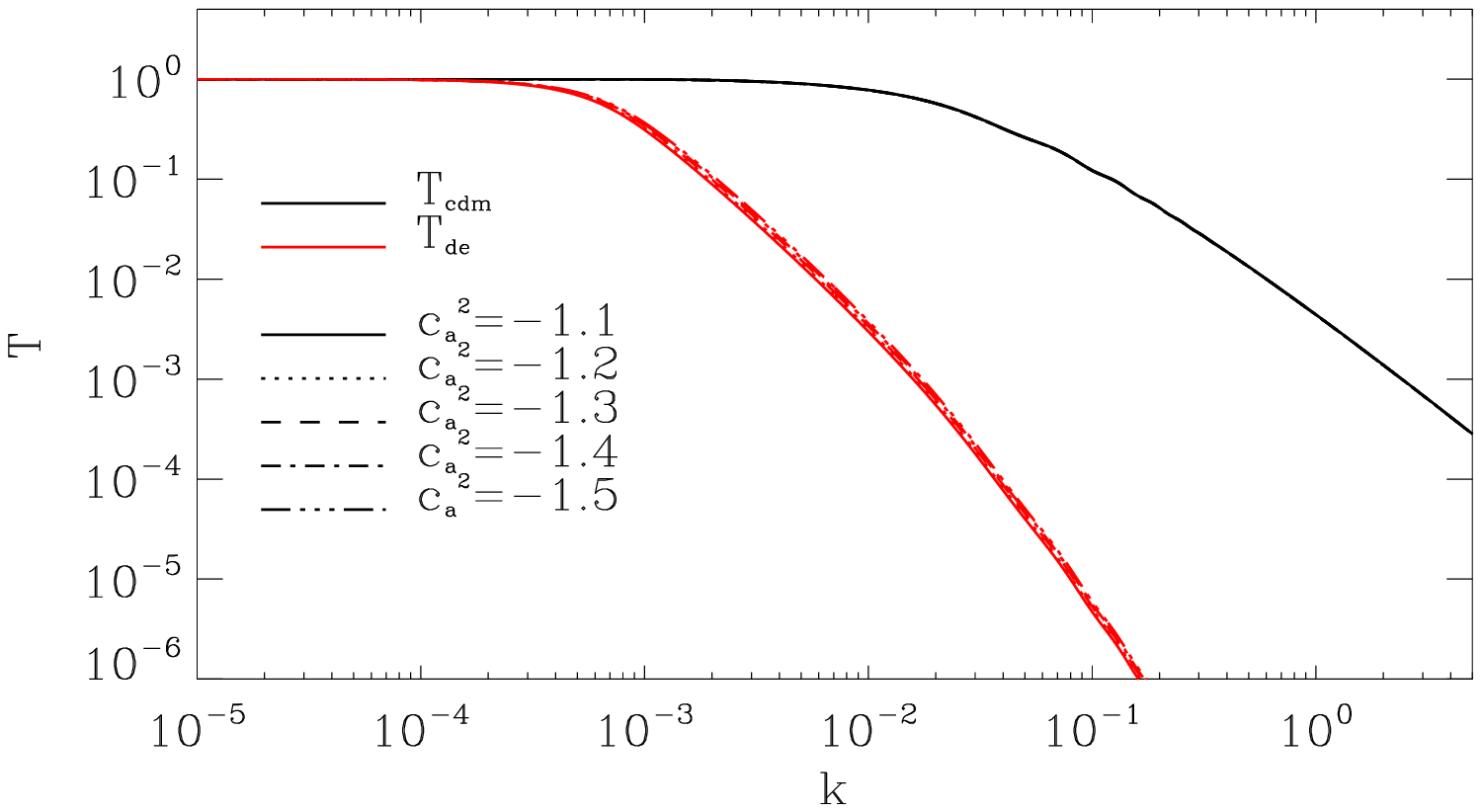}
\caption{The dark matter and dark energy transfer functions for different values of $c_s^2$ (upper panels), $w_0$ (middle) and $c_a^2$ (bottom). Left -- quintessence, right -- phantom.}
\label{trans}
\end{figure*}

Let us finally discuss the impact of effective sound speed and other dark energy parameters on the scale dependence of dark matter and dark energy perturbations. In Fig. \ref{trans} the transfer functions for dark matter and dark energy are presented for the current epoch (for discussion of dark energy transfer functions see also \cite{Sergijenko2009}). Here the perturbations for each component are calculated in its own rest frame, so the dark energy transfer functions correspond to the dark energy rest frame. In all panels the main cosmological parameters are taken from $\mathbf{p_3}$ in \cite{Novosyadlyj2014}. In the upper row the transfer functions for different values of $c_s^2$ are shown (for quintessential field here $w_0=-0.9$, $c_a^2=-0.5$, for phantom $w_0=-1.1$, $c_a^2=-1.5$). We see that the effect of value of the effective sound speed on the scale dependence of transfer function of dark energy (which is not directly observable) is quite strong, while the effect on observable scale dependence of 
cold dark matter transfer function is negligible. It is interesting to note that the scale dependence of dark energy transfer function with $c_s^2=0$ coincides with the scale dependence of dark matter transfer functions (superimposed lines in both panels) despite the sufficiently different equation of state and different frames. As it can be seen in the middle panels, the value of $w_0$ has virtually no effect on the scale dependence of both dark matter and dark energy transfer functions (here for quintessential field $c_a^2=-0.5$, $c_s^2=1$, for phantom $c_a^2=-1.5$, $c_s^2=1$). In the bottom row the scale dependences of transfer functions are shown for different values of the adiabatic sound speed. For the phantom field (right panel, $w_0=-1.1$, $c_s^2=1$) the value of $c_a^2$ has almost no influence on the scale dependence of transfer functions for both dark components. For the quintessential field (left panel, $w_0=-0.9$, $c_s^2=1$) the value $c_a^2=0$ leads to the visibly larger suppression of cold dark 
matter transfer function than other considered values, the suppression of dark energy transfer functions is different for all considered values of $c_a^2$.

\section{Observational constraints on parameters of the scalar field dark energy with $c_s^2=const$}\label{best-fit}

\subsection{Observational data and method}
To obtain joint constraints on the main cosmological parameters ($\Omega_bh^2$, $\Omega_{cdm}h^2$, $H_0$, $A_s$, $n_s$, $\tau_{rei}$) along with the independent dark energy ones for model (\ref{rhow})-(\ref{w_de}) (current value of EoS parameter $w_0$, adiabatic sound speed $c_a^2$ and effective sound speed $c_s^2$) we use the Monte Carlo Markov chain (MCMC) method implemented in the CosmoMC code \cite{cosmomc,cosmomc_source}. To compute the theory predictions we use the CAMB code assuming the Universe to be spatially flat (this allows the determination of another dark energy parameter $\Omega_{de}$) and applying for neutrinos the minimal-mass normal hierarchy of masses: a single massive eigenstate with $m_{\nu}=0.06$ eV (in accordance with Planck \cite{Planck2013c}). For the dark energy parameters we apply flat priors with ranges of values [-2,-0.33] for $w_0$, [-2,0] for $c_a^2$ and [0,1] for $c_s^2$. 

We use the following observational data:
\begin{enumerate}
\item CMB temperature fluctuations angular power spectra from Planck-2013 results \cite{Planck2013b} (together with the WMAP9 polarization \cite{wmap9a});
\item Hubble constant measurement from Hubble Space Telescope (HST) \cite{Riess2011};
\item BAO data from the galaxy surveys SDSS DR7 \cite{Padmanabhan2012}, SDSS DR9 \cite{Anderson2012}, 6dF \cite{6dF} (hereafter we quote all them together as BAO);
\item Power spectrum of galaxies from Wigglez Dark Energy Survey \cite{WiggleZ};
\item Supernovae Ia luminosity distances from either SNLS3 compilation \cite{snls3} or Union2.1 \cite{union} compilations.
\end{enumerate}
Each MCMC run has 8 chains converged to $R-1<0.01$.

We do not use here the data on CMB-LSS cross-correlation, this would be the matter of a separate paper.

\begin{table}[tbp]
\centering
 \caption{The best-fit values ($\mathbf{p}_i$), mean values and 2$\sigma$ confidence limits for parameters of
cosmological models obtained from 2 observational datasets: Planck+HST+BAO+SNLS3 ($\mathbf{p}_1$),  Planck+HST+BAO+Union2.1 ($\mathbf{p}_2$).}
  \begin{tabular}{ccccc}
    \hline\hline
    \medskip
Parameters &$\mathbf{p}_1$&mean$\pm$2$\sigma$ c.l.&$\mathbf{p}_2$&mean$\pm$2$\sigma$ c.l.\\
    \hline
\medskip
$\Omega_{de}$&0.723&0.720$_{-0.022}^{+0.021}$&0.710&0.716$_{-0.025}^{+0.023}$\\
\medskip
$w_0$&-1.176&-1.170$_{-0.134}^{+0.135}$&-1.161&-1.157$_{-0.159}^{+0.165}$\\
\medskip
$c_a^2$&-1.509&-1.373$_{-0.238}^{+0.234}$&-1.454&-1.373$_{-0.239}^{+0.246}$\\
\medskip
$c_s^2$&0.406&0.506$_{-0.506}^{+0.494}$&0.494&0.508$_{-0.508}^{+0.492}$\\
\medskip
10$\Omega_{b}h^2$&0.221&0.221$_{-0.005}^{+0.005}$&0.220&0.221$_{-0.005}^{+0.005}$\\
\medskip
$\Omega_{cdm}h^2$&0.119&0.120$_{-0.004}^{+0.004}$&0.121&0.120$_{-0.004}^{+0.004}$\\
\medskip
$h$&0.715&0.714$_{-0.027}^{+0.027}$&0.704&0.710$_{-0.031}^{+0.030}$\\
\medskip
$n_s$&0.962&0.960$_{-0.012}^{+0.012}$&0.957&0.960$_{-0.012}^{+0.012}$\\
\medskip
$\log(10^{10}A_s)$&3.095&3.089$_{-0.046}^{+0.051}$&3.089&3.089$_{-0.047}^{+0.051}$\\
\medskip
$\tau_{rei}$&0.093&0.089$_{-0.025}^{+0.026}$&0.087&0.089$_{-0.024}^{+0.026}$\\
    \hline\hline
  \end{tabular}
  \label{tab_wpbs}
\end{table}

\begin{figure*}[tbp]
\includegraphics[width=0.49\textwidth]{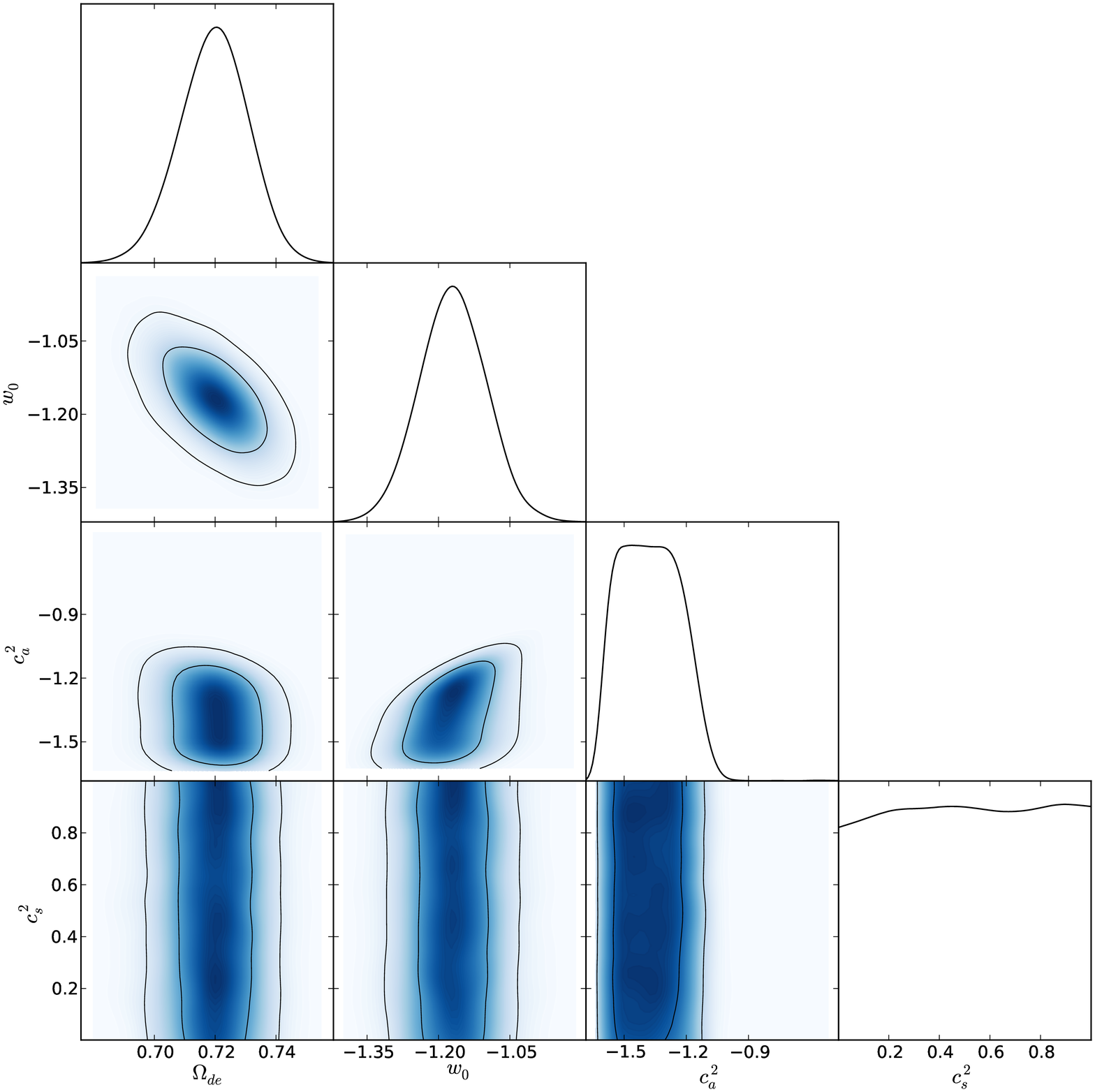}
\includegraphics[width=0.49\textwidth]{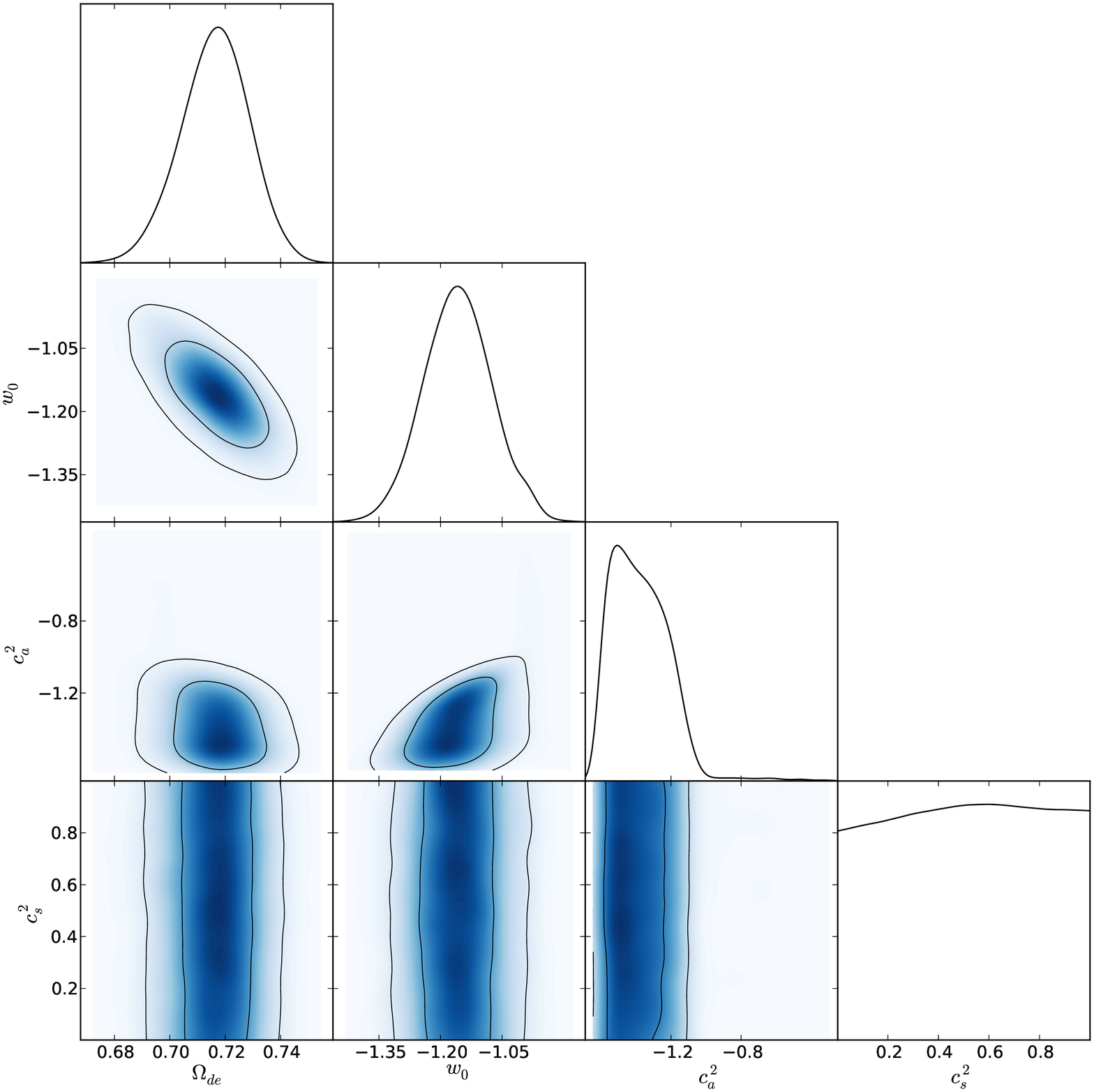}
\caption{One-dimensional marginalized posteriors (solid lines, correspond to the parameters marked below each column) for $\Omega_{de}$, $w_0$ and $c_a^2$; color panels show two-dimensional marginalized posterior distributions, where solid lines show the $1\sigma$ and $2\sigma$ confidence contours. The plots are for Planck+HST+BAO with SNLS3 (left) and Union2.1 (right) SNe Ia compilations.}
\label{postlike_wpbs}
\end{figure*}

\begin{table}[tbp]
\centering
 \caption{The best-fit values ($\mathbf{p}_i$), mean values and 2$\sigma$ confidence limits for parameters of
cosmological models obtained from 2 observational datasets: Planck+HST+WiggleZ+\\SNLS3 ($\mathbf{p}_3$), Planck+HST+WiggleZ+Union2.1 ($\mathbf{p}_4$).}
  \begin{tabular}{ccccc}
    \hline\hline
    \medskip
Parameters &$\mathbf{p}_3$&mean$\pm$2$\sigma$ c.l.&$\mathbf{p}_4$&mean$\pm$2$\sigma$ c.l.\\
    \hline
\medskip
$\Omega_{de}$&0.715&0.714$_{-0.026}^{+0.024}$&0.719&0.709$_{-0.028}^{+0.025}$\\
\medskip
$w_0$&-1.169&-1.128$_{-0.126}^{+0.134}$&-1.137&-1.111$_{-0.141}^{+0.139}$\\
\medskip
$c_a^2$&-1.380&-1.327$_{-0.253}^{+0.244}$&-1.263&-1.323$_{-0.246}^{+0.245}$\\
\medskip
$c_s^2$&0.769&0.509$_{-0.509}^{+0.491}$&0.632&0.506$_{-0.506}^{+0.494}$\\
\medskip
10$\Omega_{b}h^2$&0.219&0.221$_{-0.005}^{+0.005}$&0.222&0.221$_{-0.005}^{+0.005}$\\
\medskip
$\Omega_{cdm}h^2$&0.121&0.119$_{-0.004}^{+0.004}$&0.119&0.120$_{-0.004}^{+0.005}$\\
\medskip
$h$&0.709&0.706$_{-0.028}^{+0.028}$&0.712&0.700$_{-0.030}^{+0.030}$\\
\medskip
$n_s$&0.963&0.960$_{-0.013}^{+0.013}$&0.964&0.960$_{-0.013}^{+0.013}$\\
\medskip
$\log(10^{10}A_s)$&3.102&3.087$_{-0.047}^{+0.052}$&3.091&3.086$_{-0.046}^{+0.050}$\\
\medskip
$\tau_{rei}$&0.095&0.089$_{-0.024}^{+0.027}$&0.090&0.089$_{-0.024}^{+0.026}$\\
    \hline\hline
  \end{tabular}
  \label{tab_wwpbs}
\end{table}

\begin{figure*}[tbp]
\includegraphics[width=0.49\textwidth]{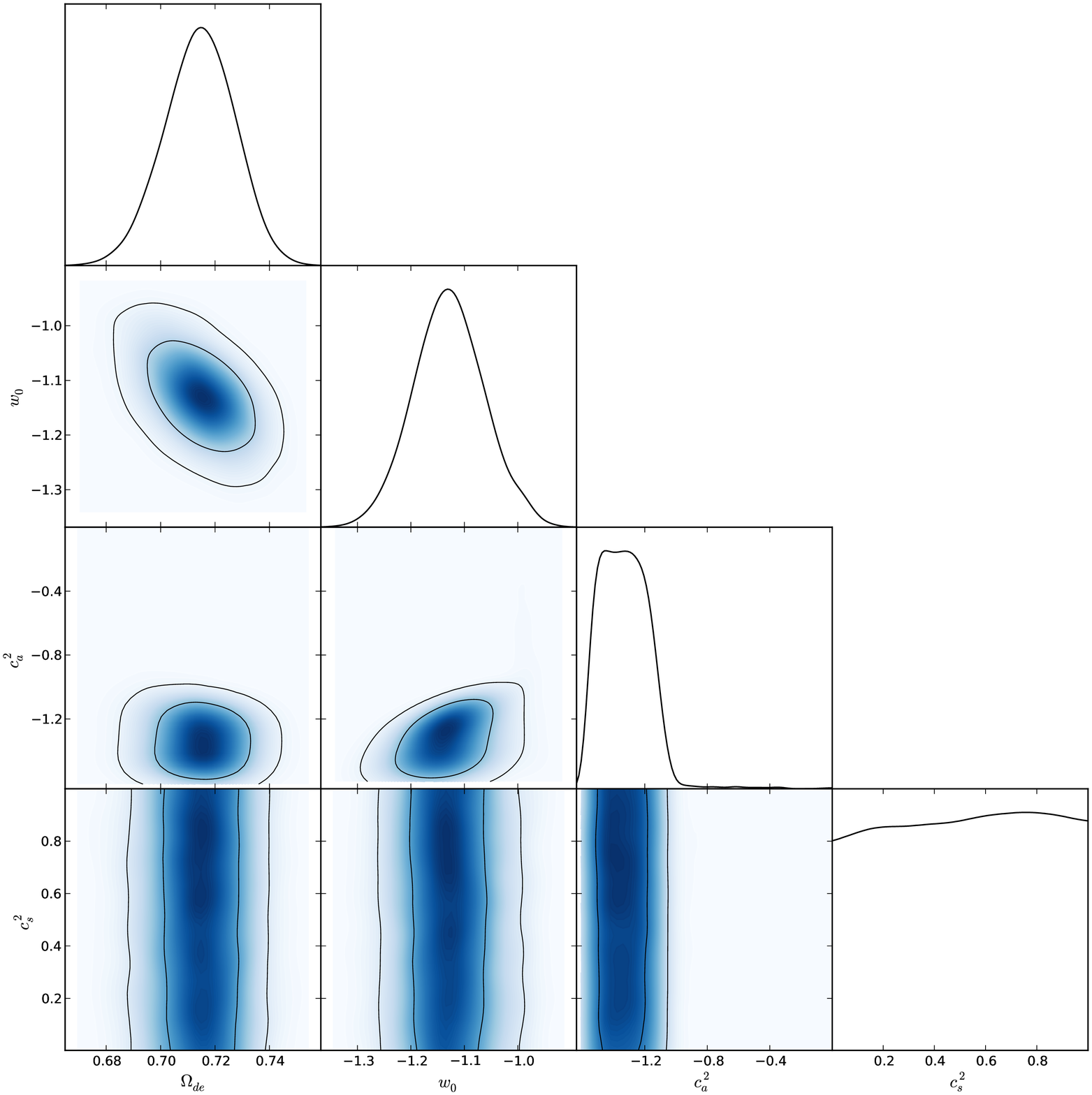}
\includegraphics[width=0.49\textwidth]{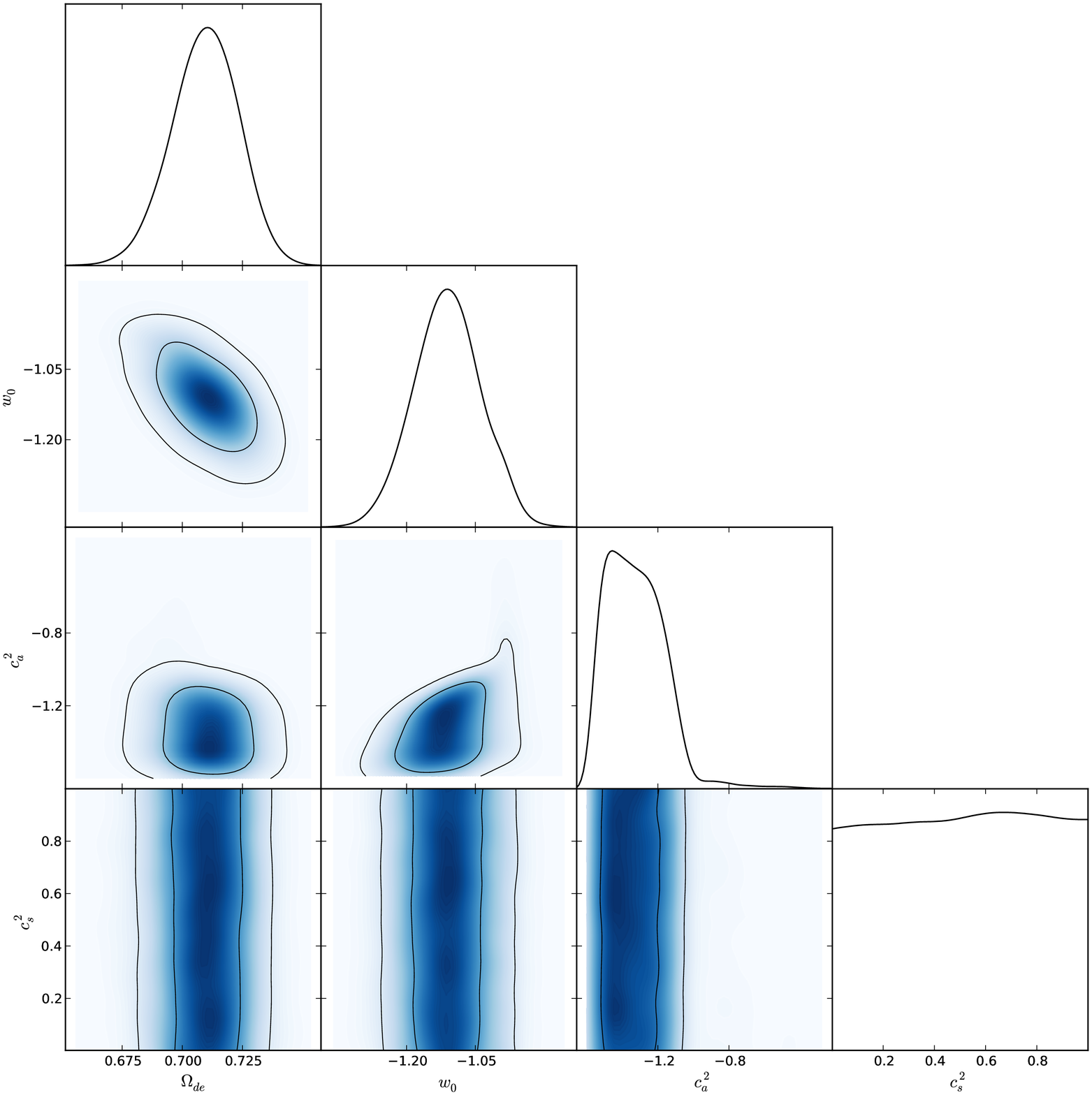}
\caption{One-dimensional marginalized posteriors (solid lines, correspond to the parameters marked below each column) for $\Omega_{de}$, $w_0$ and $c_a^2$; color panels show two-dimensional marginalized posterior distributions, where solid lines show the $1\sigma$ and $2\sigma$ confidence contours. The plots are for Planck+HST+WiggleZ with SNLS3 (left) and Union2.1 (right) SNe Ia compilations.}
\label{postlike_wwpbs}
\end{figure*}

\subsection{Results and discussion}
The obtained results are presented in Fig. \ref{postlike_wpbs} and Table \ref{tab_wpbs} for combined datasets Planck+HST+BAO+SNLS3, Planck+HST+BAO+Union2.1 and in Fig. \ref{postlike_wwpbs} and Table \ref{tab_wwpbs} for other 2 combined datasets Planck+HST+WiggleZ+SNLS3, Planck+HST+WiggleZ+Union2.1.

First of all we would like to note the progress in accuracy of determination of the cosmological parameters and parameters of the scalar field dark energy $\Omega_{de}$, $w_{de}$ and $c_a^2$, in particular as the result of enlarging of amount and quality of the observational data published in last few years. This follows from comparison of the confidential ranges in Tab. \ref{tab_wpbs} and contours in Fig. \ref{postlike_wpbs} with corresponding ones in \cite{Novosyadlyj2012} (Fig. 8, Tab. I, II) and in \cite{Novosyadlyj2013} (Tab. 1) obtained approximately 2 years ago. Now the accuracy of determination of dark energy density at current epoch reaches ~3\% at 2$\sigma$ C.L. Two years ago we had no hope of obtaining the closed contour in $w_0$, which we now have. The presented here marginalized one-dimensional posteriors for evolutionary parameter\footnote{If we expand (\ref{rhow}) into Taylor series around $a=1$ then we obtain the relation between evolutionary parameter of EoS $w_a=-3(1+w_0)(w_0-c_a^2)$ in 
the CPL approximation \cite{CPL1,CPL2} and our $c_a^2$.} of EoS $c_a^2$ are approximately Gaussian, however, they are still far from being as good as that for $w_0$.

We see in Fig. \ref{postlike_wpbs} and Tab. \ref{tab_wpbs} that the value of $c_s^2$ is unconstrained by the data: for both sets including BAO the one-dimensional posteriors for it are virtually flat, the 2$\sigma$ ranges cover the full prior range. This could be expected and agrees with the conclusions of other authors. The datasets including the power spectrum from WiggleZ were expected to provide better constraints on the effective sound speed of dark energy than those including BAO. However, as we see in Fig. \ref{postlike_wwpbs} and Tab. \ref{tab_wwpbs}, in this case the value of $c_s^2$ is also unconstrained. The precision of determination of other parameters is comparable for datasets including BAO and WiggleZ.

It is important to check whether the free value of effective sound speed affects the possibility to constrain other cosmological parameters and especially the dark energy ones. If we compare the presented in Table \ref{tab_wpbs} mean values and 2$\sigma$ limits for all parameters except for $c_s^2$ with the corresponding mean values and 2$\sigma$ limits from Table 2 of \cite{Novosyadlyj2014}, we see the good coincidence between them. So, we conclude that the problem with determination of value of $c_s^2$ has no effect on the precision and reliability of determination of values of other cosmological parameters. The differences between the best fit values (and their deviations from the mean values) come from the fact that presented best-fits correspond to the single sample with the highest likelihood obtained when generating chains, not to the result of direct minimization of $\chi^2$. It is known that MCMC method generally does not provide the accurate best-fit (global minimum of $\chi^2$), so the best-fits 
are presented here for illustration, while the mean values and marginalized limits should be regarded as the constraints. However, general conclusions about the properties of best-fit models for datasets Planck+HST+BAO+SNLS3 and Planck+HST+BAO+Union2.1 from \cite{Novosyadlyj2014} remain valid in the case of free $c_s^2$.

\begin{figure*}[tbp]
\includegraphics[width=0.49\textwidth]{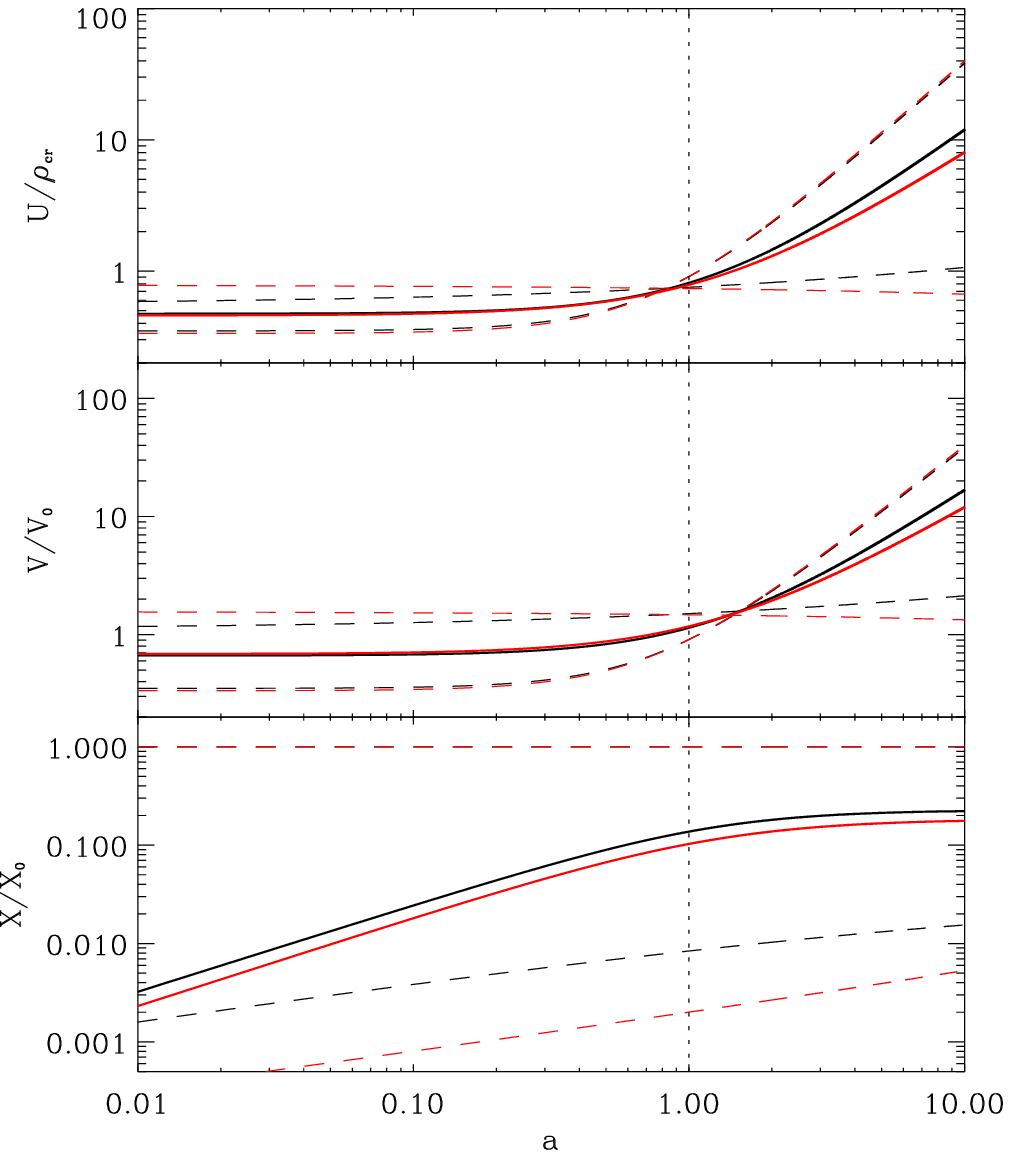}
\includegraphics[width=0.49\textwidth]{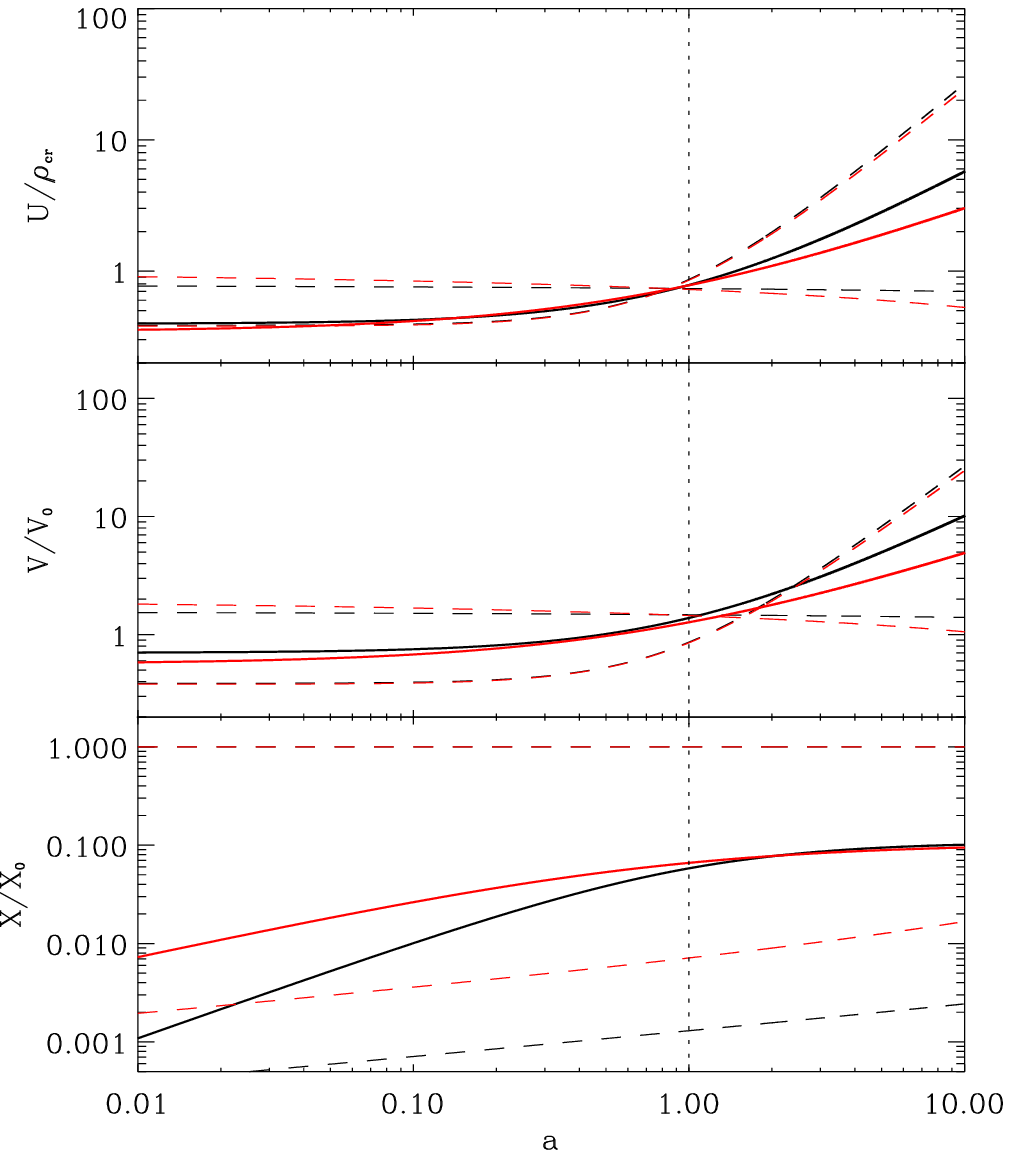}
\caption{The potentials and kinetic terms for models corresponding to the upper and lower 2$\sigma$ limits of parameters from Tables \ref{tab_wpbs} and \ref{tab_wwpbs} (dashed lines). Left column: black -- Planck+HST+BAO+SNLS3, red -- Planck+HST+BAO+Union2.1. Right column: black -- Planck+HST+WiggleZ+SNLS3, red -- Planck+HST+WiggleZ+Union2.1. The solid lines show for comparison the potentials and kinetic terms for the corresponding best fit sample models. Vertical dotted lines mark the current epoch.}
\label{uxvbf}
\end{figure*}

Finally, to make some guess about the possible uncertainties of the reconstructed scalar field Lagrangian in Fig. \ref{uxvbf} we present the potentials and kinetic terms for models corresponding to the upper and lower 2$\sigma$ limits of parameters obtained here. We see that the weakest constraints (estimated as the area between dashed lines) are, as it can be expected from (\ref{U})-(\ref{VX}) for the unconstrained $c_s^2$, for the kinetic term $X$.

\section{Conclusion}\label{conclusions}

We have analyzed the possibility of reconstruction of Lagrangian of the scalar field with $c_s^2=const$ and found that it is not unambiguous and is only possible up to an arbitrary constant. Considering the dark energy in the Universe to be the scalar field with generalized linear barotropic EoS (\ref{rhow}) and constant effective sound speed we have found that the influence of the value of $c_s^2$ on observable quantities is too weak to allow any reliable observational constraints on this parameter. Estimating the value of $c_s^2$ together with other dark energy and cosmological parameters on the basis of datasets Planck+HST+BAO+SNLS3 and Planck+HST+BAO+Union2.1 we have found that the effective sound speed remains unconstrained by these datasets while the constraints on other parameters are in good agreement with those obtained from the same datasets for the classical scalar field ($c_s^2=1$) in \cite{Novosyadlyj2014}. The datasets Planck+HST+WiggleZ+SNLS3 and Planck+HST+WiggleZ+Union2.1 also do not 
constrain the value of $c_s^2$.

\begin{acknowledgments}
This work was supported by the project of Ministry of Education and Science of Ukraine (state registration number 
0113U003059) and research program ``Scientific cosmic research'' of the National Academy of Sciences of Ukraine (state registration number 0113U002301). Authors also acknowledge the usage of CAMB, CAMB sources and CosmoMC packages.
\end{acknowledgments}

\end{document}